\documentclass[aps,prb,reprint,groupedaddress]{revtex4-1}

\usepackage{amsmath,amssymb,bm}
\usepackage{graphicx}
\usepackage{hyperref}
\usepackage{float}
\usepackage{braket}
\usepackage{color}
\usepackage{enumitem}
\usepackage{here}
\usepackage{natbib}

\usepackage{caption}
\usepackage{subcaption}
\usepackage{tikz}
\usetikzlibrary{decorations.pathreplacing}

\makeatletter
\newsavebox{\@brx}
\newcommand{\llangle}[1][]{\savebox{\@brx}{\(\m@th{#1\langle}\)}%
  \mathopen{\copy\@brx\kern-0.5\wd\@brx\usebox{\@brx}}}
\newcommand{\rrangle}[1][]{\savebox{\@brx}{\(\m@th{#1\rangle}\)}%
  \mathclose{\copy\@brx\kern-0.5\wd\@brx\usebox{\@brx}}}
\makeatother

\graphicspath{
{./},{./figure/}
}

\begin{document}

\title{Squeezed ensemble for systems with first-order phase transitions}
\author{Yasushi Yoneta}
\email{yoneta@as.c.u-tokyo.ac.jp}
\author{Akira Shimizu}
\email{shmz@as.c.u-tokyo.ac.jp}
\affiliation{Komaba Institute for Science,
 The University of Tokyo, 3-8-1 Komaba, Meguro, Tokyo 153-8902, Japan}
\affiliation{Department of Basic Science,
 The University of Tokyo, 3-8-1 Komaba, Meguro, Tokyo 153-8902, Japan}

\date{\today} 

\begin{abstract}
All ensembles of statistical mechanics are equivalent in the sense
 that they give the equivalent thermodynamic functions in the thermodynamic limit.
However, when investigating microscopic structures in the first-order phase transition region,
 one must choose an appropriate statistical ensemble.
The appropriate choice is particularly important
 when one investigates finite systems,
 for which even the equivalence of ensembles does not hold.
We propose a class of statistical ensembles,
 which
 always give the correct equilibrium state
 even in the first-order phase transition region.
We derive various formulas for this class of ensembles,
 including the one by which temperature
 is obtained directly from energy without knowing entropy.
Moreover, these ensembles are convenient for practical calculations
 because of good analytic properties.
We also derive formulas which relate
 statistical-mechanical quantities of different ensembles,
 including the conventional ones,
 for finite systems.
The formulas are useful for obtaining results
 with smaller finite size effects,
 and for improving the computational efficiency.
The advantages of the squeezed ensembles are confirmed by applying them
 to the Heisenberg model and the frustrated Ising model.
\end{abstract}

\maketitle


\section{Introduction}\label{sec:introduction}
Using statistical mechanics,
 one can obtain not only the thermodynamic functions but also the density operator,
 by which one can investigate microscopic structures of equilibrium states\cite{Gibbs1902,Landau1980}.
To obtain thermodynamic functions in the thermodynamic limit,
 one can employ any statistical ensemble
 because all statistical ensembles give the equivalent thermodynamic functions, i.e.,
 the functions are Legendre transformations of each other\cite{Ruelle1999}.
This useful property, called the {\em equivalence of ensembles},
 holds even for the system which undergoes the first-order phase transition,
 where the thermodynamic functions exhibit the strongest singularities.
By contrast,
 to investigate microscopic structures
 in the first-order phase transition region,
 one must choose an appropriate statistical ensemble.

For example, consider the liquid-gas phase transition of water
 at a pressure of $1\ \mathrm{atm}$.
If one uses the canonical ensemble specified by temperature $T$,
 the phase transition takes place at a single point
 $T=100\ {}^\circ\mathrm{C}$.
However, at this phase transition point,
 the molar ratio of the liquid and the gas phases can take various values\cite{Gibbs1902}.
Consequently, the equilibrium state changes discontinuously in temperature,
 and it is impossible to obtain equilibrium states
 for each molar ratio of  liquid and gas phases using the canonical ensemble.
By contrast, if one employs the microcanonical ensemble specified by energy,
 the phase transition takes place in a finite region of energy,
 called the {\em phase transition region} or
 {\em coexisting region}\cite{Shimizu2008,Kastner2009,Alder1962}.
At every point in this region, the ensemble
 gives the correct equilibrium state,
 in which all macroscopic variables including the liquid-gas molar ratio are uniquely determined.
When gradually heating up water in experiments
 one obtains a sequence of such equilibrium states in the transition region.

As seen from this example,
 one must choose an appropriate statistical ensemble, such as the microcanonical ensemble,
 to obtain a correct density operator and thereby investigate microscopic structures
 in the first-order phase transition region.

The appropriate choice of the ensemble
 is particularly
 important when one investigates finite systems
 because even the equivalence of ensembles (that holds in the thermodynamic limit) does not hold for finite systems
 \cite{Stump1987,Gross2001}.
That is, even when one is only interested
 in thermodynamic functions,
 one must choose an appropriate statistical ensemble
 in order to derive correct properties of finite systems from the functions.
For example, even with short-range interactions,
 finite systems which undergo first-order phase transitions
 exhibit thermodynamic anomalies such as a negative specific heat
 \cite{Bixon1989,Labastie1990,Gross1990,Gross1997}.
In fact, the recent technical development enabled
 the experimental realization of first-order phase transitions in small systems,
 and evidence of the negative specific heat was observed
 \cite{DAgostino2000,Schmidt2001}.
Nevertheless, the canonical ensemble always gives positive specific heat.
Moreover, it gives double peaks of the energy distribution\cite{Janke1998},
i.e., an unphysical state which is a classical mixture of macroscopically distinct states
\cite{Penrose1971,Binder1980}.
To correctly obtain the negative specific heat and 
a physical equilibrium state,
 one must use another ensemble such as the microcanonical ensemble
 \cite{Stump1987,Gross2001,Junghans2006,Junghans2008,Chen2009,Penrose1971,Binder1980,Troster2012}.

%

The use of an appropriate ensemble is also important
 for numerical studies of the systems which undergo the first-order phase transition.
However, conventional ensembles are not appropriate enough.
If one employs the canonical ensemble its energy distribution has double peaks,
 which are separated by exponentially suppressed phase coexisting states,
 near the transition point\cite{Schierz2016}.
This degrades greatly the efficiency of the Monte Carlo calculations
 using the importance sampling with local update algorithms
 or
 the replica exchange method \cite{Hukushima1996}
 (also called parallel tempering) \cite{Martin-Mayor2007,Kim2010}.
%
Furthermore,
 the singularities of thermodynamic quantities at the first-order transition point
 are smeared significantly
 in the canonical ensemble for finite systems due to the large fluctuation
 \cite{Stump1987,Huller1992,Huller1994}.
This makes it difficult to identify the phase transition and to determine its order
 \cite{Huller1994}.
By contrast, if such finite systems are studied using the microcanonical ensemble,
 the phase transitions are directly detected
 \cite{Challa1988_PRA,Challa1988_PRL,Behringer2005,Behringer2006}.
%
Unfortunately, however, the microcanonical ensemble has technical difficulties in
practical calculations.
It is difficult, especially for quantum systems, to construct a microcanonical ensemble.
Moreover,
 one needs to differentiate the entropy in order to calculate the temperature,
 but it gives very noisy results in numerical calculation\cite{Kanki2005}.

Several attempts were made to overcome these problems.
For example,
 the Gaussian ensemble \cite{Hetherington1987,Challa1988_PRL,Challa1988_PRA,Johal2003}
 and
 the dynamical ensemble \cite{Gerling1993}
 were conceived as elaborate numerical methods for classical systems.
Furthermore, the generalized canonical ensemble \cite{Costeniuc2005,Costeniuc2006,Toral2006} was introduced,
 which gives the entropy in the thermodynamic limit
 via the Legendre transformation
 even when the entropy is not concave.
While its mathematical aspects 
were studied,
 the physical aspects were not discussed,
 such as the physical properties of the state
 described by that ensemble.

In this paper, we propose a class of statistical ensembles,
 which we call the {\em squeezed ensembles}.
They always give the correct equilibrium state
 even in the first-order phase transition region.
In particular,
 thermodynamic anomalies, such as negative specific heat,
 are correctly obtained,
 which appear generally in the transition region
 for finite systems with short-range interactions.
We derive various formulas for this class of ensembles,
 including the one by which temperature is obtained
 directly from energy without knowing entropy.

Moreover, the squeezed ensembles are convenient for practical calculations
 because of good analytic properties.
They can be numerically constructed
 more easily than the microcanonical ensemble,
 and the construction is even easier than that of the canonical ensemble in some cases.
Furthermore, 
 efficient numerical methods, such as the replica exchange method,
 are applicable in almost the same manner as in the canonical ensemble.

We also derive formulas which relate
 statistical-mechanical quantities of different ensembles,
 including the conventional ones,
 for finite systems.
The formulas are useful for obtaining results
 with smaller finite size effects,
 and for improving the computational efficiency.
The advantages of the squeezed ensembles and these formulas are confirmed
 by applying them to
 the Heisenberg model and the frustrated Ising model.

Various ensembles,
 including the Gaussian and the dynamical ensembles of the previous works,
 are included in the class of squeezed ensembles.
One can choose an appropriate squeezed ensemble
 depending on the purpose, without losing the above advantages.
By contrast,
 the conventional ensembles,
 such as the canonical and microcanonical ensembles,
 are understood as certain limiting cases of the squeezed ensembles
 so that some of the advantages are lost by the limiting procedure.

\section{Squeezed ensemble} \label{sec:SE}
\subsection{Definition}
We consider a quantum system which has $N$ degrees of freedom and the Hamiltonian $\hat{H}$.
To take the thermodynamic limit, we use $\hat{h} \equiv \hat{H}/N$
 and the energy density
 $u \equiv \mathrm{energy}/N$ \footnote{Precisely speaking,
 $u$ is the mean energy.
It agrees with the energy density
 only when a single uniform phase is realized in the equilibrium state.
For simplicity, we use the term ``energy density'' 
 throughout this paper
 even when several phases coexist.}.
We assume that all quantities are nondimensionalized with an appropriate scale.
We denote the minimum and the maximum eigenvalues of $\hat{h}$
 by $\epsilon^{\mathrm{min}}$ and $\epsilon^{\mathrm{max}}$, respectively.

We assume that the equilibrium state
 is specified by the energy density for each value of $N$.
In other words, we assume that
 the state described by the microcanonical ensemble
 is not a classical mixture of macroscopically distinct states.

We also assume that the system is consistent with thermodynamics in the sense that
\begin{align}
  \sigma_N(u) \equiv \frac{1}{N} \log g_N(u) \label{eq:log_g}
\end{align}
converges to an $N$-independent concave function (entropy density, $s$) as $N \to \infty$,
 where $g_N(u)$ denotes the density of microstates\footnote{
More precisely, $\sigma_N$ is a twice continuously differentiable function
 which closely approximates $\frac{1}{N} \log g_N$
 and satisfies
\begin{align}
  \lim_{N\to\infty} \sigma_N^{(n)} (u) = s^{(n)}(u). \qquad (n=0,1,2)
\end{align}
We assume the existence of such $\sigma_N$.
The validity of analysis based on this assumption will be checked numerically
 in Section~\ref{sec:Heisenberg}.}.
For the moment,
 we assume that $\sigma_N$ is also a concave function for finite $N$.
Later on, it turns out that this assumption is unnecessary.
Our formulation is valid even for systems whose concavity of $\sigma_N$ is broken.

We introduce the squeezed ensemble.
Let $\eta$ be a convex function on $\left[\epsilon^{\mathrm{min}},\epsilon^{\mathrm{max}}\right]$.
We define the {\em squeezed ensemble associated with $\eta$} by
\begin{align}
  \hat{\rho}_N^\eta \equiv \frac{e^{- N \eta(\hat{h})}}{\Phi_N^\eta},
\end{align}
 where
\begin{align}
  \Phi_N^\eta \equiv \mathrm{Tr} \left[ e^{- N \eta(\hat{h})} \right].
\end{align}

When $\eta(u) = \beta u$, $\hat{\rho}_N^\eta$ gives the canonical ensemble.
When $\eta(u)= \left( \frac{u-\varepsilon}{\delta} \right)^{2n}$,
 $\hat{\rho}_N^\eta$ approaches the microcanonical ensembles as $n \to \infty$.
We will show that other, appropriate forms of $\eta(u)$ give better ensembles.

\subsection{Requirements on $\eta$}
As discussed in Section~\ref{sec:introduction},
 the canonical ensemble gives an unphysical state
 at the first-order phase transition point.
To make the squeezed ensembles free from such deficiency,
 we require
\begin{enumerate}[label={(\Alph*)}]
  \setcounter{enumi}{0}
  \item $\eta$ is a strongly convex function. \label{cond:A}
\end{enumerate}
To calculate temperature easily (using Eq.~(\ref{eq:temperature}) below)
 and to simplify the analysis, we also assume
\begin{enumerate}[label={(\Alph*)}]
  \setcounter{enumi}{1}
  \item $\eta$ is a twice continuously differentiable function. \label{cond:B}
\end{enumerate}
These conditions ensure that
 the squeezed ensembles give the correct equilibrium state
 even in the first-order phase transition region, as follows.

\begin{figure}[h]
  \centering
  \includegraphics[width=1.0\linewidth]{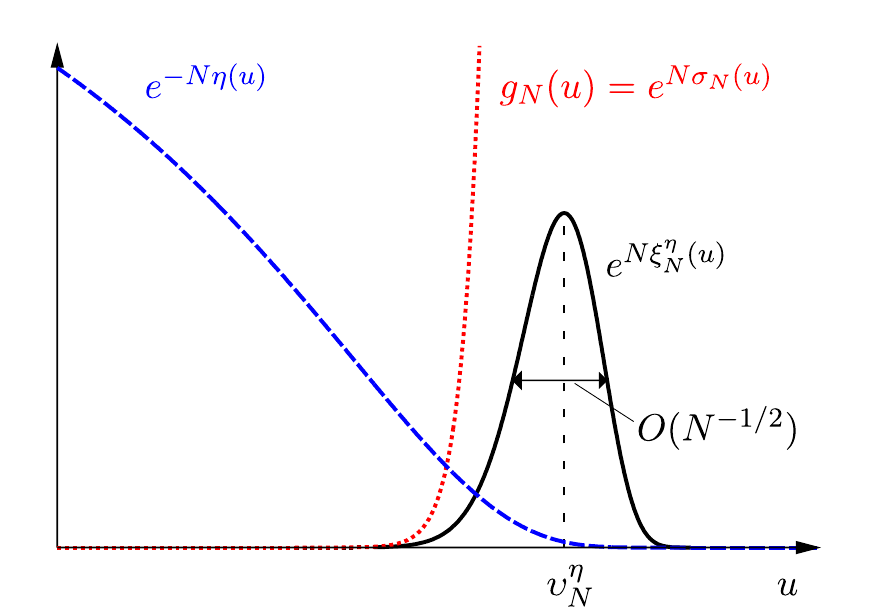}
  \caption{Schematic plot of
 the energy density distribution in $\hat{\rho}_N^\eta$, 
 which is given by $e^{N \xi_N^\eta(u)} = g_N(u) \text{(red dotted line)} \times e^{-N \eta(u)} \text{(blue dashed line)}$.}
  \label{fig:energy_dist}
\end{figure}

We examine how the energy density distributes in $\hat{\rho}_N^\eta$.
Let $f$ be an $N$-independent function.
Then
\begin{align}
  \mathrm{Tr} \left[ f(\hat{h}) e^{-N\eta(\hat{h})} \right]
  &= \int du f(u) e^{N \xi_N^\eta(u)}, \label{eq:tr_f-rho}
\end{align}
where $\xi_N^\eta(u) \equiv \sigma_N(u) - \eta(u)$.
$\xi_N^\eta(u)$ takes the maximum at $u=\upsilon_N^\eta$ which satisfies
\begin{align}
  \beta_N(\upsilon_N^\eta) \equiv \sigma_N'(\upsilon_N^\eta) = \eta'(\upsilon_N^\eta).
\label{eq:upsilon}
\end{align}
Expanding $\xi_N^\eta$ around $\upsilon_N^\eta$ and noting ${\xi_N^\eta}''(\upsilon_N^\eta)<0$, we get
\begin{align}
  \xi_N^\eta(u) = \xi_N^\eta(\upsilon_N^\eta)
  - &\frac{1}{2} \left| {\xi_N^\eta}''(\upsilon_N^\eta) \right| (u-\upsilon_N^\eta)^2 + \cdots. \label{eq:xi_taylor}
\end{align}
Hence, in the vicinity of $\upsilon_N^\eta$, $e^{N \xi_N^\eta}$ behaves as the Gaussian distribution,
 peaking at $\upsilon_N^\eta$, with the small variance $\frac{1}{N \left| {\xi_N^\eta}''(\upsilon_N^\eta) \right|}$
 (Fig.~\ref{fig:energy_dist}).

Unlike the canonical ensemble, $e^{N \xi_N^\eta}$ has a sharp peak even when $\sigma_N''(\upsilon_N^\eta)=0$
 in the first-order phase transition region
 thanks to the strong convexity of $\eta$, i.e.,
\begin{align}
  {\xi_N^\eta}''(\upsilon_N^\eta)
  = \sigma_N''(\upsilon_N^\eta)
  - \eta''(\upsilon_N^\eta) < 0. \label{eq:xi_2}
\end{align}
Therefore, as proven in Appendix~\ref{sec:proof_mech_var},
 $\hat{\rho}_N^\eta$ represents the equilibrium state
 specified by the energy density
\begin{align}
  u_N^\eta \equiv \mathrm{Tr} \left[ \hat{h} \hat{\rho}_N^\eta \right]
\end{align}
in the following senses:
\begin{enumerate} [label={(\roman*)}]
  \item We can obtain the expectation value of any ``mechanical variable'' $\hat{A}$
 in the microcanonical ensemble $\hat{\rho}_N^\mathrm{mic}$
 with energy density $u_N^\eta$ as \footnote{
We take the energy width of $\hat{\rho}_N^\mathrm{mic}$
 as specified in Appendix~\ref{sec:proof_mech_var}.}
\begin{align}
  \mathrm{Tr} \left[ \hat{A} \hat{\rho}_N^\mathrm{mic}(u_N^\eta) \right]
  = \mathrm{Tr} \left[ \hat{A} \hat{\rho}_N^\eta \right]
    \left( 1+O(N^{-1}) \right). \label{eq:mech_var_exp}
\end{align}
Here, by {\em mechanical variable},
 we mean a local observable (i.e., an observable on a continuous $O(N^0)$ sites),
 such as the two-point correlation functions,
or the sum of local operators,
 such as the total magnetization.
Hence, both ensembles give the same result in the thermodynamic limit,
 even in the first-order phase transition region.

  \item Any macroscopic additive observable $\hat{A}$
 has small variance as
\begin{align}
  \mathrm{Tr}\left[\hat{a}^2 \hat{\rho}_N^\eta\right]
  - \mathrm{Tr}\left[\hat{a} \hat{\rho}_N^\eta\right]^2
  = o(N^0), \label{eq:mech_var_var}
\end{align}
 where $\hat{a} \equiv \hat{A}/N$.
\end{enumerate}

Obviously, the state described by this ensemble depends on $\eta$.
In order to obtain the states specified by a series of energies,
 it is convenient for practical calculations if $\eta$ depends on a parameter.
We will discuss this parameter dependence in Section~\ref{sec:parameter}.

\subsection{Genuine thermodynamic variables} \label{sec:thermo_var}
We can also obtain genuine thermodynamic variables such as the entropy and temperature.
Using Eqs.~(\ref{eq:tr_f-rho})-(\ref{eq:xi_2}) and applying Laplace's method, we have
\begin{align}
  \frac{1}{N} \log \Phi_N^\eta &= \sigma_N(\upsilon_N^\eta) - \eta (\upsilon_N^\eta) + O(N^{-1} \log N), \label{eq:massieu}\\
  u_N^\eta &= \upsilon_N^\eta+O(N^{-1}).\label{eq:u_upsilon}
\end{align}
Therefore, we obtain
\begin{align}
  \sigma_N(u_N^\eta) = \frac{1}{N} \log \Phi_N^\eta + \eta\left(u_N^\eta\right) + O(N^{-1} \log N).
  \label{eq:entropy}
\end{align}
It is sometimes convenient to rephrased this relation as
\begin{align}
  \sigma_N(u_N^\eta) = s_N^\mathrm{vN}(\hat{\rho}_N^\eta) + O(N^{-1} \log N),
\end{align}
 where $s_N^\mathrm{vN}$ is the von Neumann entropy density,
\begin{align}
  s_N^\mathrm{vN}(\hat{\rho}) \equiv - \frac{1}{N} \mathrm{Tr} \left[ \hat{\rho} \log \hat{\rho} \right].
\end{align}
Using Eqs.~(\ref{eq:upsilon}) and (\ref{eq:u_upsilon}), we also obtain
\begin{align}
  \beta_N(u_N^\eta) = \eta'\left(u_N^\eta\right) + O(N^{-1}).
  \label{eq:temperature}
\end{align}
Here, since $\eta' = \sigma_N'$ only at $\upsilon_N^\eta$,
 replacing $\upsilon_N^\eta$ with $u_N^\eta$ yields the difference of $O(N^{-1})$.

In the thermodynamic limit, $\sigma_N$ and $\beta_N$ converge to 
the thermodynamic entropy density and inverse temperature, respectively,
 which are well-defined in the thermodynamic limit.
Therefore, one can obtain the temperature of the equilibrium state specified by $u_N^\eta$ just by calculating $u_N^\eta$,
 via Eq.~(\ref{eq:temperature}).
By contrast, in order to calculate the temperature using the microcanonical ensemble, one needs to differentiate the entropy,
 and it gives very noisy results in numerical calculation\cite{Kanki2005}.

In a similar manner, we obtain
\begin{align}
  c_N(u_N^\eta)
  &\equiv \left( \frac{d \left(1/\beta_N\right)}{du}(u_N^\eta) \right)^{-1} \nonumber\\
  &= \frac{\left(\beta_N(u_N^\eta)\right)^2}{\frac{1}{N \mathrm{Tr} \left[\left(\hat{h}-u_N^\eta\right)^2 \hat{\rho}_N^\eta \right]}-\eta''(u_N^\eta)+ O(N^{-1})}.
  \label{eq:specificheat}
\end{align}
In the thermodynamic limit, $c_N$ converges to the thermodynamic specific heat.


\subsection{Interpretation}
Although the above results have been derived naturally
 using
 Laplace's method,
 the following scenario may be more intuitive from the viewpoint of the principle of equal weight.

Let us consider the equilibrium state of the target system which is in weak thermal contact with an external system.
We note the ratio of the degrees of freedom of the external system to that of the target system.
Using the principle of equal weight,
 one can obtain the canonical ensemble as the ensemble of the target system if the external system is much larger than the target system,
 and the microcanonical ensemble if the  external system is much smaller than the target system.
Then, we now consider the case where the external system has about the same degrees of freedom as the target system.
In this case, the details of the external system affect the state of the target system
 and there are infinitely many ensembles
 as the state of the target system that converge the same equilibrium state as $N \to \infty$.

Suppose that the target system is in weak thermal contact with the external system which has the same degrees of freedom as the target system
 and that the restriction to the interval $[-\epsilon^\mathrm{max}, -\epsilon^\mathrm{min}]$
 of the $\sigma_N$ of the external system is equal to $-\eta(-u)$.
Assume that the target system plus the external system together are isolated, with fixed total energy $0$.
Then, applying the principle of equal weight to the total system,
 we get $\hat{\rho}_N^\eta$ as an ensemble of the target system.
We can extract all statistical-mechanical quantities about the target system
 because we are familiar with $\sigma_N$ of the external system.

\section{Parameter of squeezed ensemble} \label{sec:parameter}
Suppose that $\eta$ depends on a certain parameter.
The equilibrium states described by the squeezed ensembles are specified by the parameter.
We now examine the parameter dependencies of 
the physical quantities defined by the squeezed ensemble.

\subsection{Energy density} \label{sec:parameterdependence}
Let $K$ be a real interval for the parameter $\kappa$,
 and $\eta$ be a function on $K \times [\epsilon^\mathrm{min},\epsilon^\mathrm{max}]$.
Assume that $\eta(\kappa,\cdot)$ satisfies conditions~\ref{cond:A}-\ref{cond:B} for all $\kappa$.
Then, for each $\kappa$, $\eta(\kappa,\cdot)$ corresponds to a squeezed ensemble.
Thus, a quantity related to the squeezed ensemble can be regarded as a function of $\kappa$.
To examine the parameter dependence, we assume
\begin{enumerate}[label={(\Alph*)}]
  \setcounter{enumi}{2}
  \item $\eta$ is a twice continuously differentiable function of two variables. \label{cond:C}
\end{enumerate}

For treating the systems which undergo the first-order phase transition,
 it is necessary that
 the energy density $u_N^\eta(\kappa)$ in the squeezed ensemble
 takes every possible value of the energy density of the system.
The canonical ensemble does not satisfy this condition
 because $\kappa$ of the canonical ensemble
 corresponds to the inverse temperature $\beta$.
In the thermodynamic limit, 
 $u_N^\mathrm{can}(\beta)$ changes discontinuously in $\beta$
 at the first-order phase transition point.
Hence, as discussed in Section~\ref{sec:introduction},
 the canonical ensemble is unable to describe
 the equilibrium states
 in the first-order phase transition region,
 in which the energy density takes continuous values.
%
%
%
[As we will discuss in Section~\ref{sec:non-concave}, the correct state is not obtained even if the system is finite.]
By contrast, in the case of the squeezed ensemble,
 the energy density changes continuously in $\kappa$
 thanks to the strong convexity of $\eta(\kappa,\cdot)$ for each $\kappa$.
In fact,
\begin{align}
  \frac{\partial \upsilon_N^\eta}{\partial \kappa}(\kappa)
  &= \frac{\displaystyle
    \frac{\partial^2 \eta}{\partial \kappa \partial u}(\kappa,\upsilon_N^\eta(\kappa))
  }{\displaystyle
    \sigma_N''(\upsilon_N^\eta(\kappa))-\frac{\partial^2 \eta}{\partial u^2}(\kappa, \upsilon_N^\eta(\kappa))
  } \label{eq:upsilon_p}
\end{align}
 is finite even in the first-order phase transition region.
Furthermore, we take $\eta$ such that
\begin{enumerate}[label={(\Alph*)}]
  \setcounter{enumi}{3}
  \item \hfill
        \vspace{-\abovedisplayskip}
        \vspace{-\baselineskip}
        \begin{align}
          \inf_\kappa u_N^\eta(\kappa) = \epsilon^\mathrm{min}, \qquad
          \sup_\kappa u_N^\eta(\kappa) = \bar{\epsilon}, \nonumber
        \end{align} \label{cond:D}
\end{enumerate}
 where $\bar{\epsilon}$ is the arithmetic mean of the eigenvalues of $\hat{h}$.
Then, $u_N^\eta(\kappa)$ takes every possible value of $u$
 in the physical region
 $\epsilon^\mathrm{min} < u < \bar{\epsilon}$ \footnote{
We say the region $\epsilon^\mathrm{min} < u < \bar{\epsilon}$ is physical
 because temperature is positive in this region.}.


\subsection{Thermodynamic function} \label{sec:TDfunction}

As in the case of the conventional ensembles,
 we consider the logarithm of the ``partition function''
\begin{align}
  \psi_N^\eta (\kappa) \equiv -\frac{1}{N} \log \Phi_N^\eta (\kappa), \label{eq:psi}
\end{align}
 which is a function not of a physical quantity (such as $\beta$)
 but of our parameter $\kappa$.
In numerical calculations (using, e.g., the Monte Carlo calculation),
 $\psi_N^\eta (\kappa)$ can be obtained easily by integrating
\begin{align}
  \frac{\partial \psi_N^\eta}{\partial \kappa} (\kappa)
  = \mathrm{Tr} \left[ \frac{\partial \eta}{\partial \kappa}\left(\kappa,\hat{h}\right) \hat{\rho}_N^\eta(\kappa) \right].
\end{align}
Here, the right hand side is obtained simply
 by calculating the expectation value of $\displaystyle \frac{\partial \eta}{\partial \kappa}\left(\kappa,\hat{h}\right)$.

Let us define the {\em thermodynamic function associated with $\eta$} as
the thermodynamic limit of $\psi_N^\eta$: 
\begin{align}
  \psi^\eta(\kappa) \equiv \lim_{N \to \infty} \psi_N^\eta(\kappa).
\end{align}
As proven in Appendix~\ref{sec:proof_equivalence},
 $\psi^\eta$ is equivalent to the thermodynamic entropy density in the following sense:
\begin{align}
s(u) = \inf_\kappa \left\{ \eta(\kappa, u)-\psi^\eta(\kappa) \right\}. 
\label{eq:psi-s}
\end{align}
Using this relation, 
 one can obtain the thermodynamic entropy from $\psi^\eta$ without knowing $u_N^\eta$.
We can also invert this relation as
\begin{align}
\psi^\eta(\kappa) = \inf_u \left\{ \eta(\kappa, u)-s(u) \right\}. 
\label{eq:s-psi}
\end{align}
From a physical point of view, these relations are 
a generalization of the equivalence of the entropy density and the canonical free energy density.
From a mathematical point of view, this is a generalization of the Legendre transformation,
 to which it reduces for $\eta(\beta,u)=\beta u$.

\begin{figure*}
  \begin{minipage}[b]{0.32\linewidth}
    \centering
    \includegraphics[keepaspectratio, width=1.05\linewidth]{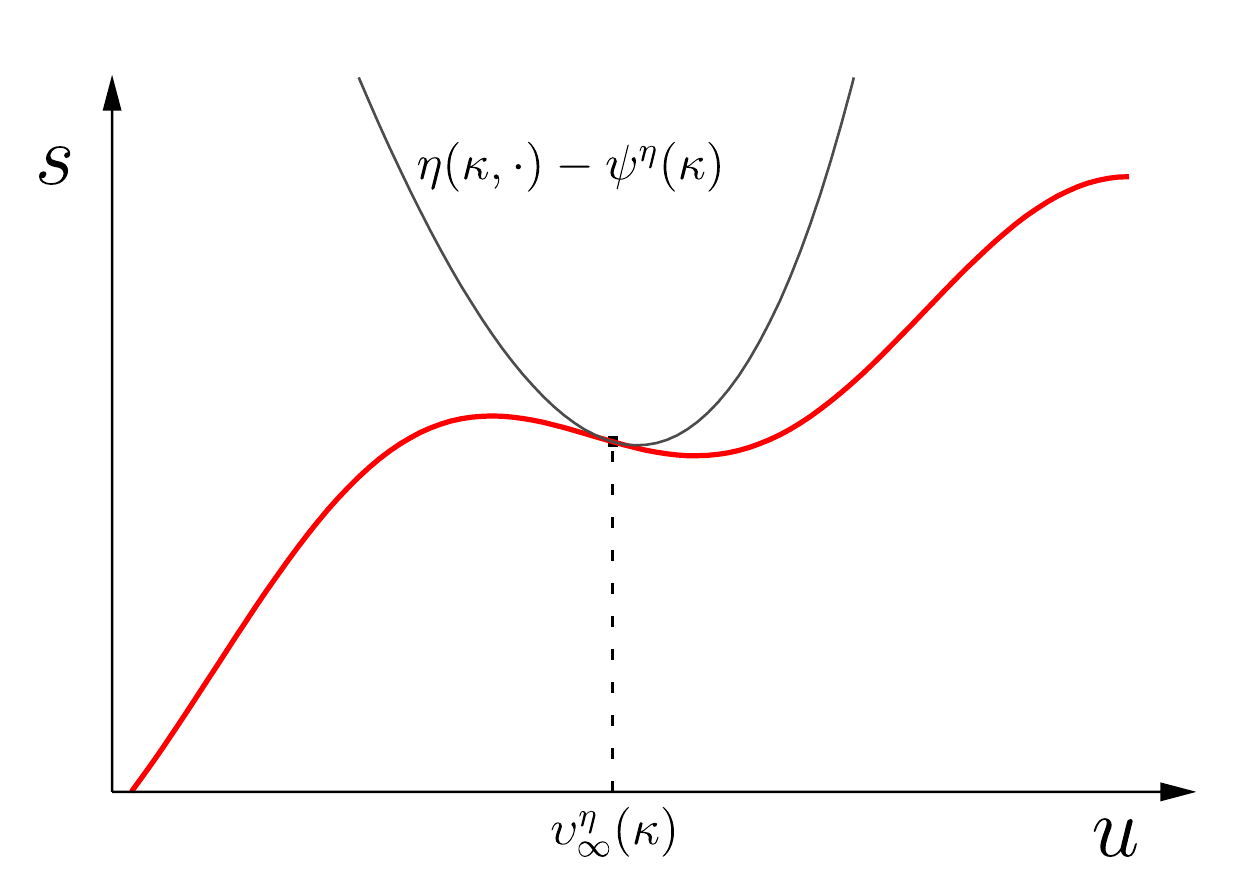}
    \subcaption{}\label{subfig:sigma-eta}
  \end{minipage}
  \begin{minipage}[b]{0.32\linewidth}
    \centering
    \includegraphics[keepaspectratio, width=1.05\linewidth]{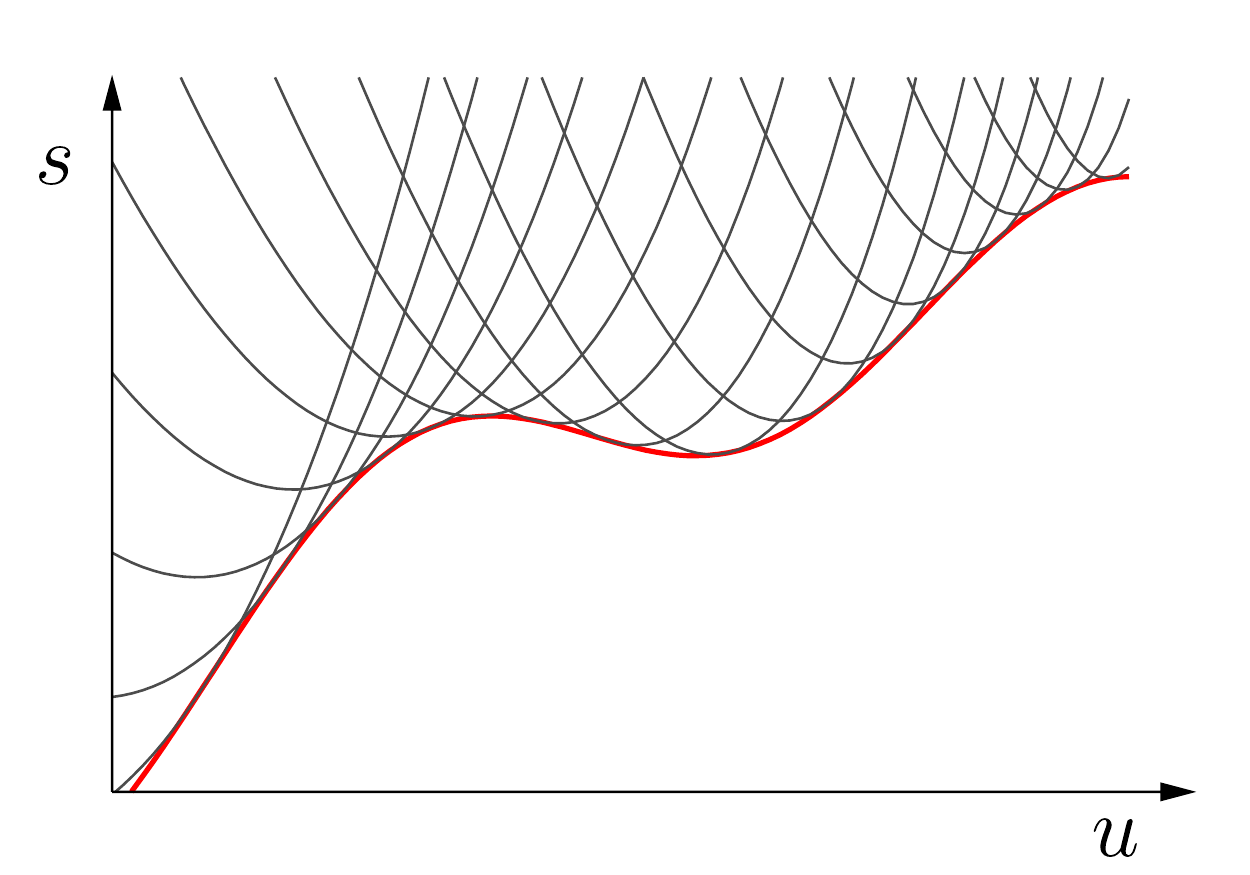}
    \subcaption{}\label{subfig:genLegendre}
  \end{minipage}
  \begin{minipage}[b]{0.32\linewidth}
    \centering
    \includegraphics[keepaspectratio, width=1.05\linewidth]{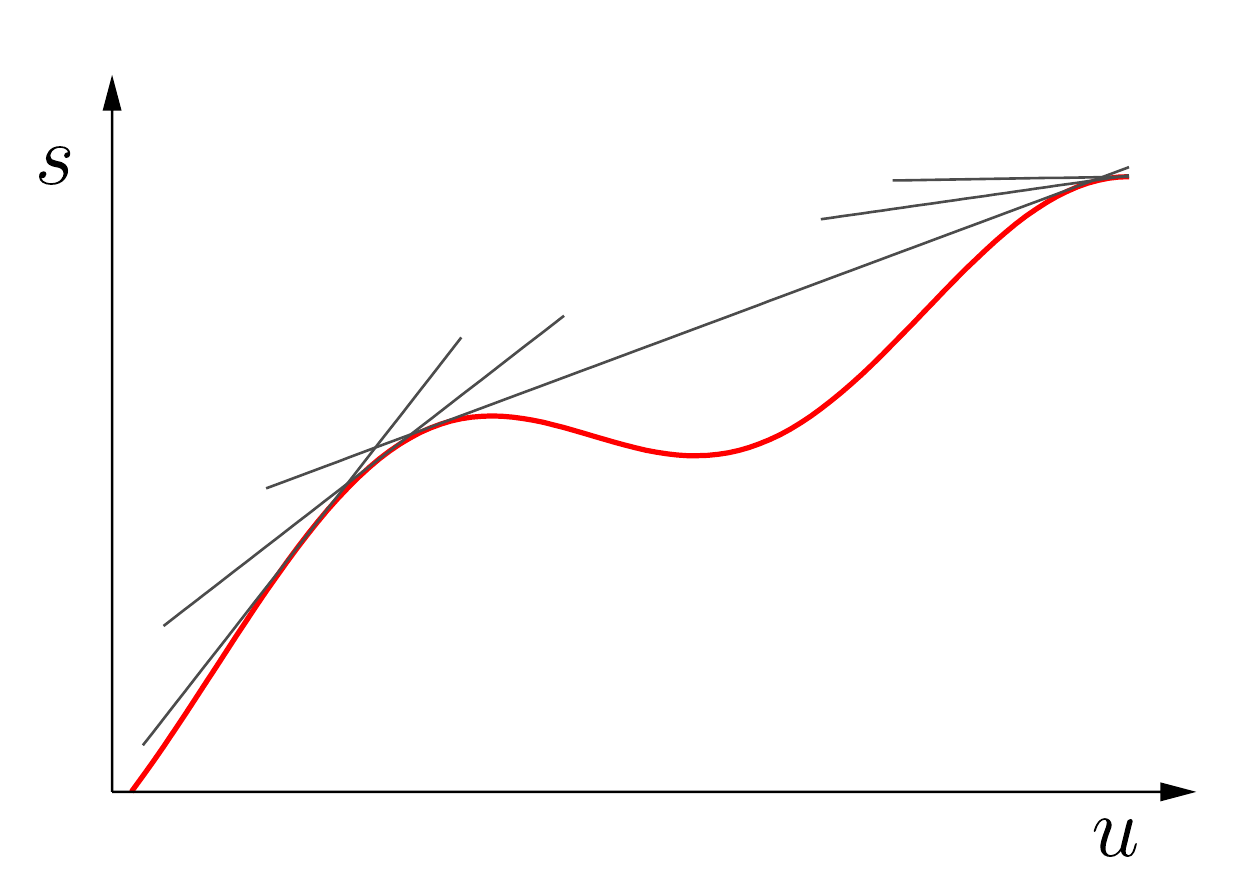}
    \subcaption{}\label{subfig:Legendre}
  \end{minipage}
  \caption{
Schematic representations of
 Eq.~(\ref{eq:psi-s})-(\ref{eq:s-psi})
 and the Legendre transformation,
for the case where the concavity of $s$ is broken.
\subref{subfig:sigma-eta} For all $u$, there exists $\kappa$ such that $\upsilon_\infty^\eta(\kappa)=u$,
 at which $\eta(\kappa, \cdot)-\psi^\eta(\kappa)$ is tangent to $s$.
\subref{subfig:genLegendre} We can reconstruct $s$ from the sets of the curves which is characterized by $\psi^\eta$.
\subref{subfig:Legendre} The Legendre transformation does not preserve the information in the non-convex function.
The set of the lines forms the convex hull of $s$.}
  \label{fig:Legendre}
\end{figure*}

It is worth mentioning that
 Eqs.~(\ref{eq:psi-s})-(\ref{eq:s-psi}) are valid
 even when the concavity of the entropy is broken,
 as long as conditions~\ref{cond:F}-\ref{cond:G} of Section~\ref{sec:non-concave}
 are satisfied.
Even in such a case, 
 we can obtain entropy from $\psi^\eta$, 
as schematically explained in Fig.~\ref{subfig:sigma-eta}-\subref{subfig:genLegendre}.
By contrast, the Legendre transformation does not preserve the information
 in the non-convex function (Fig.~\ref{subfig:Legendre}).

\section{Non-concave $\sigma_N$} \label{sec:non-concave}
So far we have assumed that $\sigma_N$ is concave.
Since we consider systems with short-range interactions,
 this assumption is valid in the thermodynamic limit.
However, for finite $N$,
 it was suggested that the concavity of $\sigma_N$ is broken
 in a system which undergoes a first-order phase transition
 with a phase separation\cite{Bixon1989,Labastie1990,Gross1990,Gross1997},
 as illustrated schematically in Fig.~\ref{subfig:nc_sigma}.
In Appendix~\ref{sec:concav_breaking}, we prove this under reasonable conditions.



\begin{figure*}
  \begin{minipage}[b]{0.32\linewidth}
    \centering
    \includegraphics[keepaspectratio, width=\linewidth]{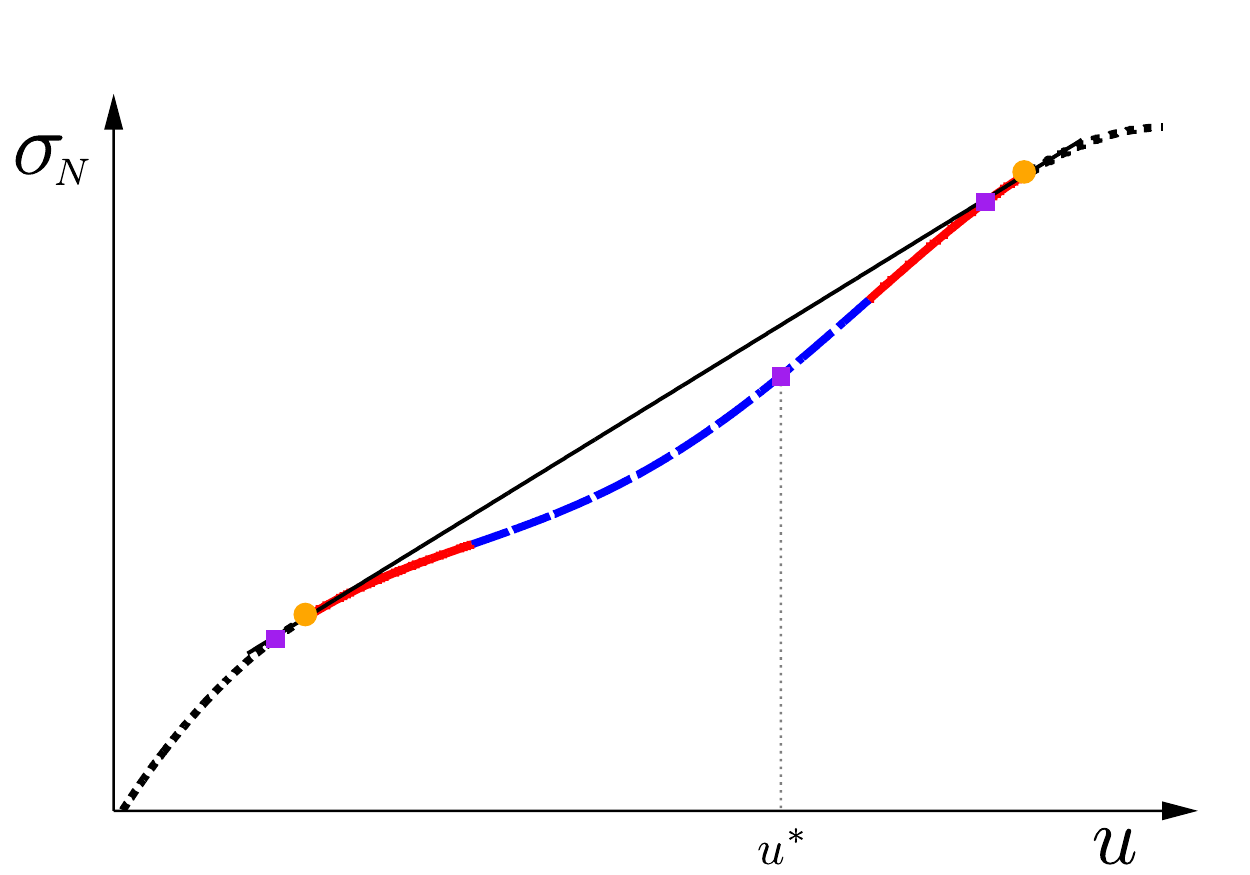}
    \subcaption{}\label{subfig:nc_sigma}
  \end{minipage}
  \begin{minipage}[b]{0.32\linewidth}
    \centering
    \includegraphics[keepaspectratio, width=\linewidth]{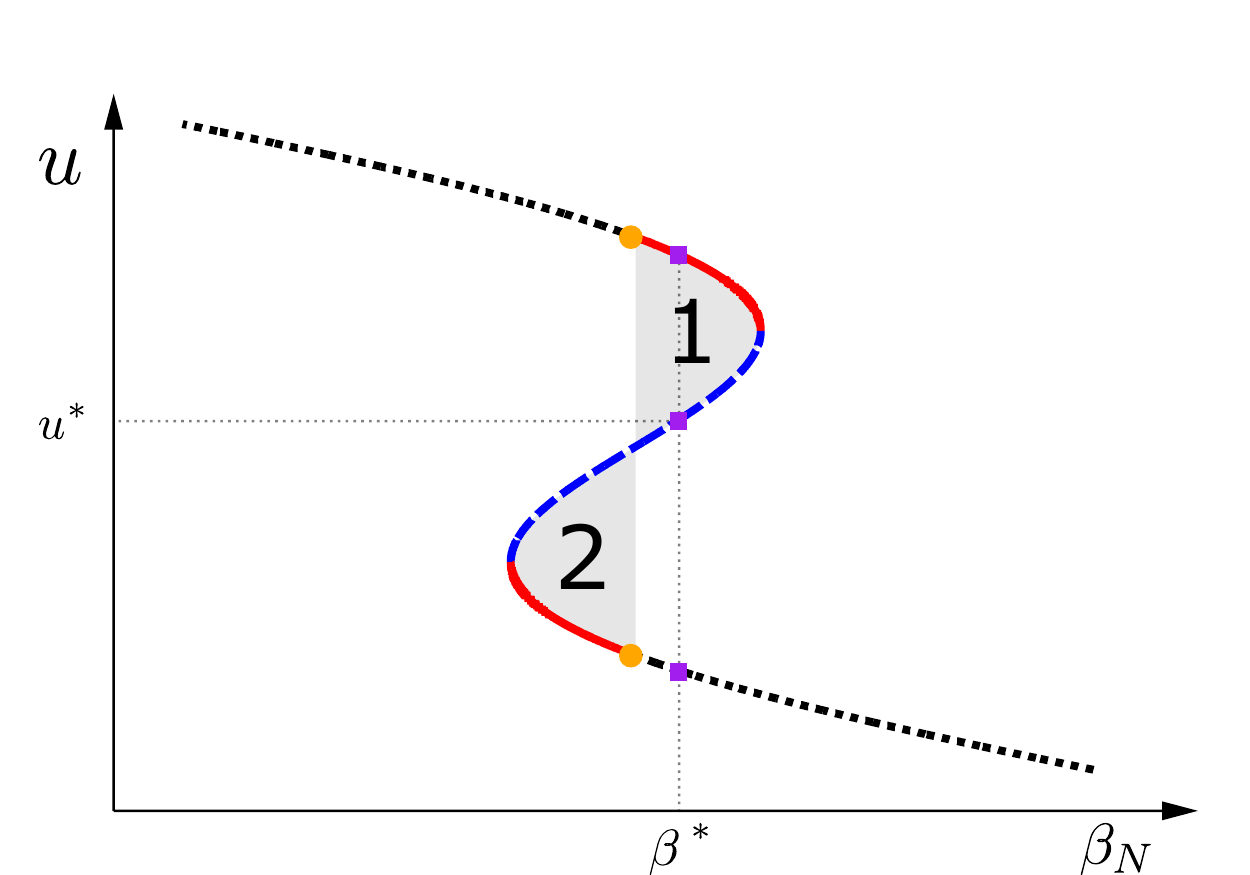}
    \subcaption{}\label{subfig:nc_beta}
  \end{minipage}
  \begin{minipage}[b]{0.32\linewidth}
    \centering
    \includegraphics[keepaspectratio, width=\linewidth]{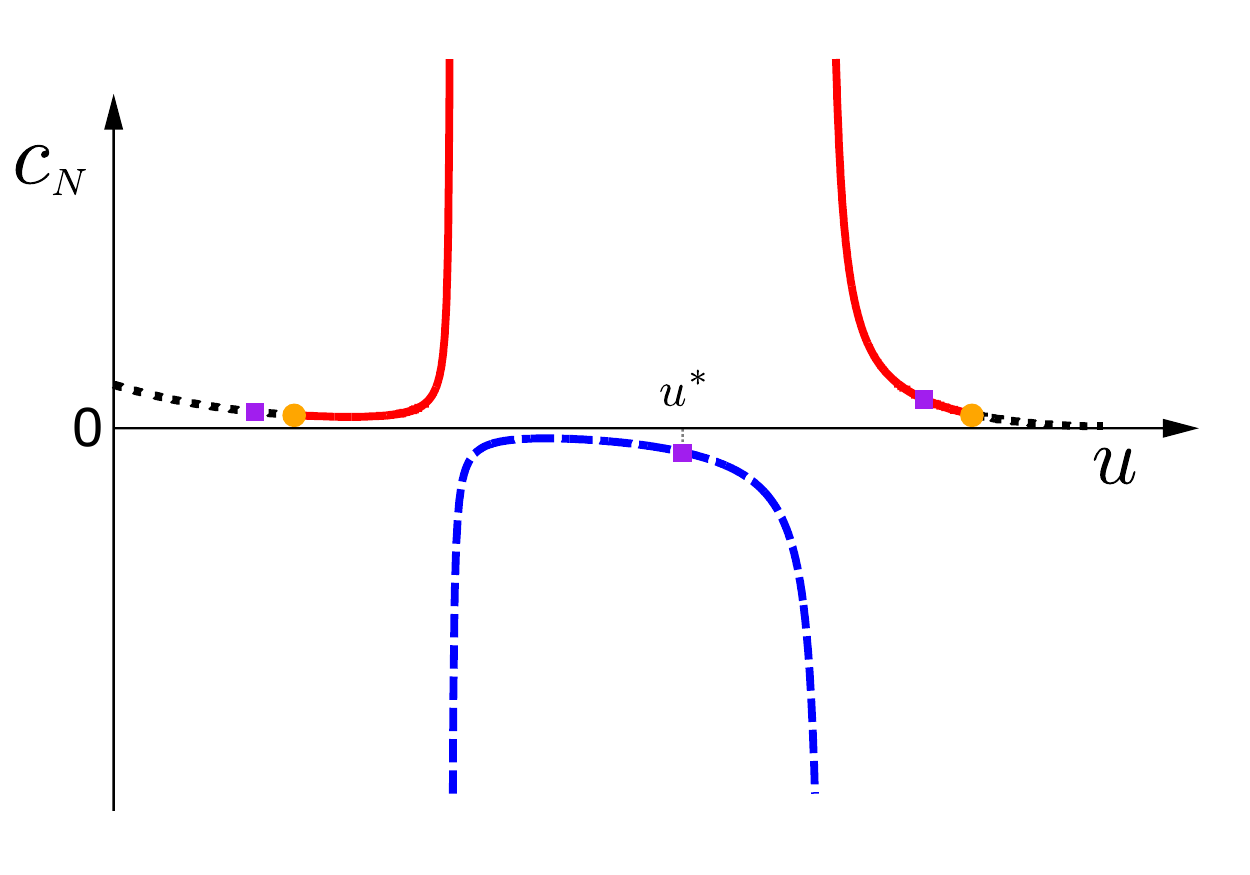}
    \subcaption{}\label{subfig:nc_c}
  \end{minipage}
  \caption{Schematic diagrams of
  \subref{subfig:nc_sigma} non-concave $\sigma_N$,
  \subref{subfig:nc_beta} $\beta_N$, and
  \subref{subfig:nc_c} $c_N$.
  In \subref{subfig:nc_sigma},
  the boundaries (orange circles) between the black dotted and the red solid line
  are intersections between $\sigma_N$ and the double tangent line.
  These points correspond exactly to the points
  in \subref{subfig:nc_beta}
  obtained using the equal area law
  (i.e., areas 1 and 2 are equal).}
  \label{fig:non-concave}
\end{figure*}

In order to obtain the correct equilibrium state even in such a case,
 we require
\begin{enumerate}[label={(\Alph*)}]
  \setcounter{enumi}{4}
  \item $\sigma_N-\eta$ has a single peak \label{cond:F}
\end{enumerate}
 and
\begin{enumerate}[label={(\Alph*)}]
  \setcounter{enumi}{5}
  \item $\sigma_N''-\eta''$ is negative at the peak. \label{cond:G}
\end{enumerate}
In fact, under these conditions,
 the energy density distribution in $\hat{\rho}_N^\eta$
 can still be well approximated by the Gaussian distribution,
 and all results in Sections~\ref{sec:SE}-\ref{sec:parameter} are valid.

Therefore, even for a system whose $\sigma_N$ is not concave,
 we can obtain the equilibrium states and the statistical-mechanical quantities
 using a squeezed ensemble associated with an appropriately chosen $\eta$.
By contrast, using the canonical ensemble,
 the equilibrium states
 with the energy density
 in the region other than the black dotted lines in Fig.~\ref{fig:non-concave}
 cannot be obtained
 (see Appendix~\ref{sec:non-concave_canonical}).

\section{Physical natures of a system with non-concave $\sigma_N$} \label{sec:phys_nat_non-concave}
The concavity breaking of $\sigma_N$ causes thermodynamic anomalies.
In this section,
 we discuss physical natures of equilibrium states with non-concave $\sigma_N$.

\subsection{Realization of equilibrium states with non-concave $\sigma_N$}
In general, an isolated system with sufficiently complex dynamics
 evolves into (relaxes to) an equilibrium state.
During this ``thermalization'' process,
 the energy density keeps the same value as that of the initial state,
 which may be a non-equilibrium state such as a local equilibrium state.
In this way,
 one can obtain equilibrium states of any possible value of the energy density
 by tuning the energy density of the initial state.

In particular, one can obtain an equilibrium state
 in the region where concavity of $\sigma_N$ is broken
 by tuning the energy density of the initial state in such a region.
%
Such an experiment will be possible in various finite systems, such as cold atoms.

A more practical way is to heat the system slower enough
 (across the transition temperature)
 so that a quasistatic process is realized.

\subsection{Measurability of statistical-mechanical quantities}

As illustrated schematically in Fig.~\ref{subfig:nc_beta}, 
$\beta_N$ has an S-shaped curve.
It is a physical quantity that can be measured as follows.

Let place the target system in weak thermal contact with an external system.
We assume the external system is sufficiently small
 so that its effect on the target system is negligible.
That is, the external system works as a thermometer and the target system as a heat bath.
After the thermal equilibrium of the total system is reached,
 the thermometer is in the canonical Gibbs state with $\beta=\beta_N(u)$,
 where $u$ is the energy density of the target system,
 because the canonical typicality \cite{Sugita2006,Sugita2007,Popescu2006,Goldstein2006}
 holds regardless of the concavity of $\sigma_N$ of the heat bath.
Hence, one can read the value of $\beta_N(u)$ from the thermometer.
Using this setup with various values of $u$,
 one can measure the function $\beta_N$ and $c_N$
 of the target system.
Then one will find not only the S-shaped $\beta_N$
 but also the anomalous behaviors of $c_N$.
The latter is illustrated schematically in Fig.~\ref{subfig:nc_c},
 where $c_N$ takes negative values between the two singular points.
See Refs.~\cite{Stodolsky1995,Chomaz1999,Schmidt1997} for related discussions.

Evidence of the concavity breaking of $\sigma_N$ due to the first-order phase transition
 has been observed both in experiments\cite{DAgostino2000,Schmidt2001}
 and in dynamical simulations\cite{Litz1992,Nielsen1994,Reyes-Nava2003}.

\subsection{Thermodynamic stability}
Even when $\sigma_N$ is not concave the equilibrium states of an isolated finite system is stable
 because of the energy conservations.
In this subsection, we discuss whether the equilibrium states are stable
 when the system is in thermal contact with an infinitely large heat bath.

Suppose that the system is initially isolated
 and in the equilibrium state with energy density $u$.
Then let the system, which we now call the target system,
 interact with a large heat bath of inverse temperature $\beta$ that is equal to $\beta_N(u)$.
According to the canonical typicality\cite{Sugita2006,Sugita2007,Popescu2006,Goldstein2006},
 the target system is driven by the heat bath toward the canonical Gibbs state.
However, as discussed in Appendix~\ref{sec:non-concave_canonical},
 if $u$ lies in the red solid or blue dashed region of Fig.~\ref{fig:non-concave} the equilibrium state of the target system
 is not realized as a canonical Gibbs state.
Therefore, for such $u$,
 the equilibrium state of the target system is no longer stable
 when attached to a heat bath.

More specifically, if $u$ lies in the blue dashed region of Fig.~\ref{fig:non-concave}
 the target system quickly evolves into the canonical Gibbs state of the same temperature
 by absorbing or emitting energy from or to the heat bath.
That is, the equilibrium state for such $u$ is thermodynamically {\em unstable}.
On the other hand, if $u$ lies in the red solid region
 the equilibrium state of the target system is thermodynamically {\em metastable}.
That is, the state is maintained for a certain macroscopic time scale
 because $\sigma_N(u)$ is locally concave in such a region
 \cite{Antoni2004}.

We can thus classify three types of regions in Fig.~\ref{fig:non-concave},
 for the case where the target system is in thermal contact with a heat bath,
 as the stable equilibrium states (black dotted line),
 the metastable states (red solid line),
 and the unstable states (blue dashed line).



\section{Relation with other ensembles in finite systems} \label{sec:inequivalence}
In finite systems, different ensembles are 
 not completely equivalent.
The inequivalence is most prominent in the first-order transition region.
Let us investigate the effects of the inequivalence.

\subsection{Inequivalence of ensembles in finite systems}
As an example, we consider the entropy density.
What one can calculate using statistical mechanics is a sequence indexed by $N$ such as
\begin{align}
  s_N^\eta(\kappa) \equiv \eta(\kappa, u_N^\eta(\kappa))-\psi_N^\eta(\kappa).
\end{align}
As $N \to \infty$,
 this $s_N^\eta(\kappa)$ converges to thermodynamic entropy density $s(u)$ at $u=u_\infty^\eta(\kappa)$.
In other words, there exists $o(N^0)$ quantity $\delta s_N^\eta(\kappa)$
 such that
\begin{align}
  s(u_\infty^\eta(\kappa)) = s_N^\eta(\kappa) + \delta s_N^\eta(\kappa).
\end{align}
On the other hand, for another pair of $(\tilde{\eta}, \tilde{\kappa})$
 which satisfies $u_\infty^{\tilde{\eta}}(\tilde{\kappa}) = u_\infty^\eta(\kappa)$,
 $s_N^{\tilde{\eta}}(\tilde{\kappa})$ also converges to $s(u_\infty^\eta(\kappa))$,
 and there exists $o(N^0)$ quantity $\delta s_N^{\tilde{\eta}}(\tilde{\kappa})$
 such that $s(u_\infty^\eta) = s_N^{\tilde{\eta}}(\tilde{\kappa}) + \delta s_N^{\tilde{\eta}}(\tilde{\kappa})$.
Generally, $s_N^\eta(\kappa)$ and $s_N^{\tilde{\eta}}(\tilde{\kappa})$ are different in finite systems.
The difference is $o(N^0)$ because $s_N^\eta(\kappa)-s_N^{\tilde{\eta}}(\tilde{\kappa})=\delta s_N^\eta(\kappa)-\delta s_N^{\tilde{\eta}}(\tilde{\kappa})$.
Note that the rates at which $\delta s_N^\eta(\kappa)$ and $\delta s_N^{\tilde{\eta}}(\tilde{\kappa})$ converge to $0$
 are different in general.

The same is true for other thermodynamic quantities.
Suppose that one calculates a thermodynamic quantity of $O(N^0)$.
What one can calculate using statistical mechanics is
 a function sequence indexed by $N$ that converges to the thermodynamic quantity as $N \to \infty$.
Therefore, there is an arbitrariness in the choice of the function sequence up to $o(N^0)$.
In order to emphasize the distinction from a thermodynamic quantity,
 we call the function sequence that converges to the thermodynamic quantity a ``statistical-mechanical quantity''.

\subsection{Formulas relating different ensembles}
One might think that it is meaningless to calculate quantities with 
 such an arbitrariness accurately.
However, it is important in numerical calculation, where one has to deal with finite systems inevitably,
 for the following reasons.
The rates of convergence of the statistical-mechanical quantities to the thermodynamic quantities with increasing $N$
 are different among ensembles.
If we can choose an ensemble which converges very quickly,
 it is advantageous for practical calculations.
On the other hand, there may also be an ensemble that gives thermodynamic quantities that converge very slowly,
 even though it is equivalent to other ensembles in the thermodynamic limit.
Using the following formulas, we can evaluate the difference between
 the statistical-mechanical quantity calculated using a squeezed ensemble and
 that calculated using a conventional ensemble (such as the canonical ensemble and microcanonical ensemble).
If necessary, we can correct the difference from the conventional ensemble.

It is known that, for a system without phase transition,
 the statistical-mechanical quantities in the canonical ensemble,
 among those in various Gibbs ensembles,
 are closest to thermodynamic quantities of the infinite system\cite{Iyer2015}.
Therefore, it is useful to derive formulas by which a squeezed ensemble gives statistical-mechanical quantities in the canonical ensemble.
Such formulas are obtained as follows.

First, let us derive formulas by which a squeezed ensemble gives statistical-mechanical quantities in another squeezed ensemble.
Below in this section, we assume that $\eta$ is differentiable as many times as necessary.
Using Eqs.~(\ref{eq:tr_f-rho})-(\ref{eq:xi_taylor}) and (\ref{eq:psi}), we have
\begin{widetext}
\begin{align}
  \psi_N^\eta
  &= \eta(\upsilon_N^\eta)
  - \sigma_N(\upsilon_N^\eta)
  - \frac{1}{2N} \log \frac{2\pi}{N |{\xi_N^\eta}''(\upsilon_N^\eta)|}
  + O(N^{-2}), \label{eq:log_phi_high}\\
  u_N^\eta
  &=\upsilon_N^\eta
  + \frac{{\xi_N^\eta}'''(\upsilon_N^\eta)}{2N \left|{\xi_N^\eta}''(\upsilon_N^\eta)\right|^2}
  + O(N^{-2}) \label{eq:u_high}
\end{align}
\end{widetext}
for any choice of $\eta$ that satisfies conditions~\ref{cond:A}-\ref{cond:B}.
Suppose that two squeezed ensembles associated with $\eta$ and $\tilde{\eta}$
 satisfy $\upsilon_N^\eta = \upsilon_N^{\tilde{\eta}}$.
Then, we have
\begin{widetext}
\begin{align}
  \psi_{\tilde{\eta}}
  &=  \psi_N^\eta
  + \frac{1}{2N} \log \left| 1+\frac{\eta''(\upsilon_N^\eta)-\tilde{\eta}''(\upsilon_N^\eta)}{{\xi_N^\eta}''(\upsilon_N^\eta)} \right|
  + \tilde{\eta}(\upsilon_N^\eta) - \eta(\upsilon_N^\eta) 
  + O(N^{-2}),\\
  u_N^{\tilde{\eta}}
  &= u_N^\eta
  - \frac{{\xi_N^\eta}'''(\upsilon_N^\eta)}{2N \left|{\xi_N^\eta}''(\upsilon_N^\eta)\right|^2}
  + \frac{
    {\xi_N^\eta}'''(\upsilon_N^\eta)+\eta'''(\upsilon_N^\eta)-\tilde{\eta}'''(\upsilon_N^\eta)
  }{
    2N \left|{\xi_N^\eta}''(\upsilon_N^\eta)+\eta''(\upsilon_N^\eta)-\tilde{\eta}''(\upsilon_N^\eta)\right|^2
  } + O(N^{-2}).
\end{align}
\end{widetext}
These formulas relate statistical-mechanical quantities between different squeezed ensembles.

\subsection{Formulas relating the squeezed ensemble to the canonical ensemble}
Next, using these formulas, let us derive formulas
 by which the squeezed ensemble gives the canonical entropy density $s^\mathrm{can}_N$ and the canonical energy density $u^\mathrm{can}_N$.
Assume that $\sigma_N''(\upsilon_N^\eta) < 0$.
This assumption is satisfied in many cases except when $\upsilon_N^\eta$ is in the phase transition region.
Set $\tilde{\eta} = \beta u$, where
\begin{align}
  \beta = \eta'(\upsilon_N^\eta). \label{eq:beta_can}
\end{align}
Then, the squeezed ensemble associated with $\tilde{\eta}$ gives 
the canonical ensemble at inverse temperature $\beta$.
Using the relation
\begin{align}
  \psi_N^\mathrm{can}(\beta) = \beta u_N^\mathrm{can}(\beta) - s_N^\mathrm{can}(\beta),
\end{align}
 we get
\begin{widetext}
\begin{align}
  s^\mathrm{can}_N(\beta)
  &= \eta(\upsilon_N^\eta) - \psi_N^\eta
  - \frac{1}{2N} \log \left( 1+\frac{\eta''(\upsilon_N^\eta)}{{\xi_N^\eta}''(\upsilon_N^\eta)} \right)
  + \frac{
    \eta'(\upsilon_N^\eta) \left({\xi_N^\eta}'''(\upsilon_N^\eta)+\eta'''(\upsilon_N^\eta)\right)
  }{
    2N \left| {\xi_N^\eta}''(\upsilon_N^\eta)+\eta''(\upsilon_N^\eta) \right|^2
  } + O(N^{-2}), \label{eq:entropy_can}\\
  u^\mathrm{can}_N(\beta)
  &= \upsilon_N^\eta
  + \frac{{\xi_N^\eta}'''(\upsilon_N^\eta)+\eta'''(\upsilon_N^\eta)}{2N \left| {\xi_N^\eta}''(\upsilon_N^\eta)+\eta''(\upsilon_N^\eta) \right|^2}
  + O(N^{-2}). \label{eq:energy_can}
\end{align}
\end{widetext}
These are the desired formulas.

Though $\upsilon_N^\eta$ is unknown, using Eq.~(\ref{eq:u_high}), we find
\begin{align}
  \upsilon_N^\eta &= u_N^\eta - \frac{{\xi_N^\eta}'''(\upsilon_N^\eta)}{2N \left|{\xi_N^\eta}''(\upsilon_N^\eta)\right|^2} + O(N^{-2}) \label{eq:upsilon_high}
\end{align}
One can calculate $\xi_N^\eta$'s derivatives at $\upsilon_N^\eta$ from the central moments of energy distribution.
For example, as the second and third order derivatives, we have
\begin{align}
  {\xi_N^\eta}''(\upsilon_N^\eta)
  &= - \frac{1}{N \mathrm{Tr} \left[ \left(\hat{h} - u_N^\eta\right)^2 \hat{\rho}_N^\eta \right]} + O(N^{-1}) \label{eq:xi_pp}\\
  {\xi_N^\eta}'''(\upsilon_N^\eta)
  &=   \frac{\mathrm{Tr} \left[ \left(\hat{h} - u_N^\eta\right)^3 \hat{\rho}_N^\eta \right]}{N { \mathrm{Tr} \left[ \left(\hat{h} - u_N^\eta\right)^2 \hat{\rho}_N^\eta \right]}^3}
  + O(N^{-1}). \label{eq:xi_ppp}
\end{align}

Using Eqs.~(\ref{eq:upsilon_high})-(\ref{eq:xi_ppp}),
 we can calculate $\upsilon_N^\eta$ and $\xi_N^\eta$'s derivatives at $\upsilon_N^\eta$
 with accuracy up to $O(N^{-1})$.
Substituting these into Eqs.~(\ref{eq:beta_can})-(\ref{eq:energy_can}),
 we can calculate the statistical-mechanical quantities in the canonical ensemble
 from the statistical-mechanical quantities in the squeezed ensemble
 with accuracy up to $O(N^{-1})$.


\section{Application to Heisenberg model} \label{sec:Heisenberg}
Before studying a system which undergoes a first-order phase transition,
we confirm the validity of our results
by applying our formulation to the Heisenberg chain, defined by the Hamiltonian
\begin{align}
  \hat{H} \equiv - J \sum_{i} \hat{\mathbf{S}}_{i} \cdot \hat{\mathbf{S}}_{i+1},
\end{align}
 where $J=+1$ (ferromagnetic).
The exact results at finite temperature have been derived for $N \to \infty$\cite{Nakamura1994}.

We can freely choose $\eta$ for practical convenience according to the physical situation.
Here, we choose $K=(0, \infty)$ and $\eta(\kappa,u)=-2\kappa\log(l-u)$, which is particularly convenient for quantum systems,
 because $\hat{\rho}_N^\eta$ and $\Phi_N^\eta$ are obtained by simply multiplying the Hamiltonian repeatedly.
In this model, $-0.75 \leq \epsilon^\mathrm{min}<\bar{\epsilon}<\epsilon^\mathrm{max} \leq +0.75$.
Hence, we take $l=1(>\epsilon^\mathrm{max})$, then $\eta(\kappa,\cdot)$ satisfies conditions~\ref{cond:A}-\ref{cond:B} for all $\kappa>0$.
Furthermore, $\eta$ satisfies conditions~\ref{cond:C}-\ref{cond:D}.

\begin{figure}
  \centering
  \includegraphics[width=1.0\linewidth]{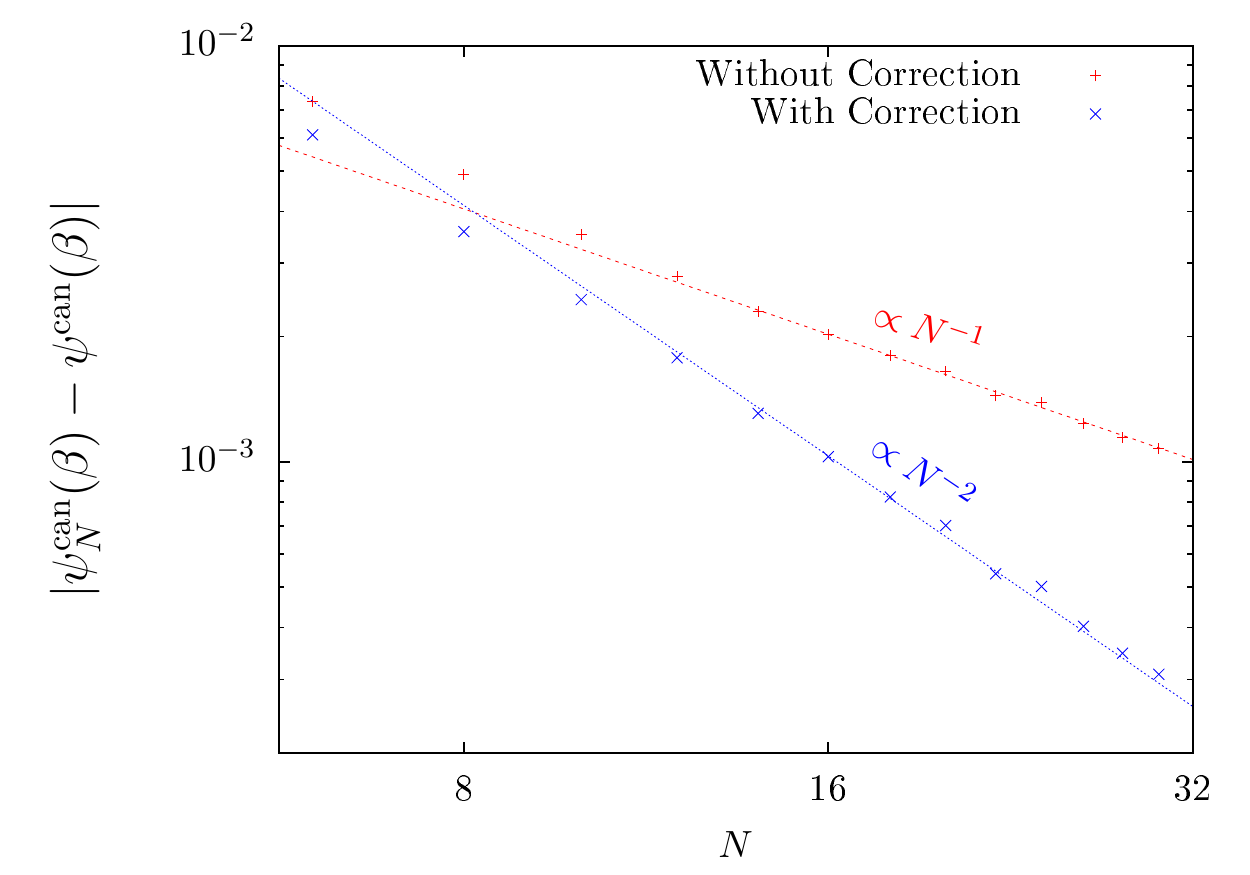}
\caption{Difference between $\psi_N^\mathrm{can}$ (obtained using the squeezed ensemble) 
 and the exact canonical free energy density of the infinite system.
$\psi_N^\mathrm{can}$ is calculated using Eq.~(\ref{eq:psiNcan}) (red crosses)
 and corrected using Eqs.~(\ref{eq:beta_can})-(\ref{eq:xi_ppp}) (blue crosses).
}
  \label{fig:Heisenberg_L-diff}
\end{figure}

We calculate statistical-mechanical quantities in the squeezed ensemble
 using thermal pure quantum formulation\cite{Sugiura2012}.
Then, using Eqs.~(\ref{eq:massieu})-(\ref{eq:temperature}),
 we calculate $\psi_N^\mathrm{can}$,
 with an error of $O(N^{-1} \log N)$,
 from the statistical-mechanical quantities in the squeezed ensemble as
\begin{widetext}
\begin{align}
  \psi_N^\mathrm{can} \left( \frac{\partial \eta}{\partial u} (\kappa, u_N^\eta) \right)
  = \psi_N^\eta(\kappa)
  + \eta(\kappa, u_N^\eta(\kappa))
  - \frac{\partial \eta}{\partial u} (\kappa, u_N^\eta) u_N^\eta(\kappa) + O(N^{-1} \log N), \label{eq:psiNcan}
\end{align}
\end{widetext}
 which corresponds to the result obtained in the previous work \cite{Sugiura2013}.
The result at $\beta \simeq 2.82$ is plotted by the red crosses in Fig.~\ref{fig:Heisenberg_L-diff}.

Furthermore, we correct the difference from the canonical free energy density
 using Eqs.~(\ref{eq:beta_can})-(\ref{eq:xi_ppp}),
 which reduces an error to $O(N^{-2})$.
The result is plotted by the blue crosses in Fig.~\ref{fig:Heisenberg_L-diff}.
It is confirmed that the difference from the canonical free energy density is proportional to $N^{-2}$.

\section{Application to Frustrated Ising model} \label{sec:FrustratedIsing}
To confirm the advantages of our formulation
 for studying the systems which undergo first-order phase transitions,
 we apply our formulation to the two-dimensional frustrated Ising model on the square lattice,
 defined by the Hamiltonian
\begin{align}
  \hat{H} \equiv - J_1 \sum_{\langle  i, j \rangle}  \hat{\sigma}^z_i \hat{\sigma}^z_j
                 + J_2 \sum_{\llangle i, j \rrangle} \hat{\sigma}^z_i \hat{\sigma}^z_j.
\end{align}
Here, $\hat{\sigma}^z_i = \pm 1$ is a classical variable, and
 $\langle i, j \rangle$ and $\llangle i, j \rrangle$ denote 
the nearest and the next-nearest neighbors, respectively.
We take both couplings $J_1, J_2$ positive, 
and measure energy density and temperature in units of $J_1$.
It is known that the model undergoes a weak first-order phase transition for $0.5 < J_2/J_1 \lesssim 0.67$\cite{Jin2012}.
We take $J_2/J_1=0.6$.
The numerical simulations are performed for systems of size $L \times L$ with periodic boundary conditions.

We choose $K=(-\infty, \bar{\epsilon})$ and $\eta(\kappa,u)=\frac{1}{2} \lambda (u-\kappa)^{2}$.
This $\eta$ satisfies conditions~\ref{cond:A}-\ref{cond:D} for positive $\lambda$
 and defines the so-called Gaussian ensemble introduced by Hetherington\cite{Hetherington1987}.
As proven in Appendix~\ref{sec:concav_breaking},
 the concavity of $\sigma_N$ is broken for finite $N$.
Since the second order derivative of $\eta(\kappa,\cdot)$ is given by $\lambda$,
 $\eta$ satisfies conditions~\ref{cond:F}-\ref{cond:G} for sufficiently large $\lambda$
 even in the case of the non-concave $\sigma_N$.
In the classical systems, it is easy to calculate
 the energy of a given configuration
 and the energy change due to a change in the local configuration.
Therefore, this choice of $\eta$ is convenient for classical systems which undergo first-order phase transitions.

We calculate the expectation values of mechanical variables in the squeezed ensemble
 using the Monte Carlo calculations.
The acceptance probability can be easily computed using the Metropolis algorithm\cite{Metropolis1953}.
As argued in Section~\ref{sec:parameter},
 the equilibrium state changes continuously in $\kappa$.
Therefore the replica exchange method\cite{Hukushima1996} works well\cite{Kim2010}
 even in the first-order phase transition region, unlike the canonical ensemble.
Other advantages of
 the Gaussian ensemble
 for studying
 the phase transitions 
 were already studied in Ref.~\cite{Challa1988_PRL,Challa1988_PRA}.
These advantages are enjoyed also by
 other choices of 
 $\eta$ as long as conditions~\ref{cond:F}-\ref{cond:G} are satisfied.

\begin{figure}
  \centering
  \includegraphics[width=1.0\linewidth]{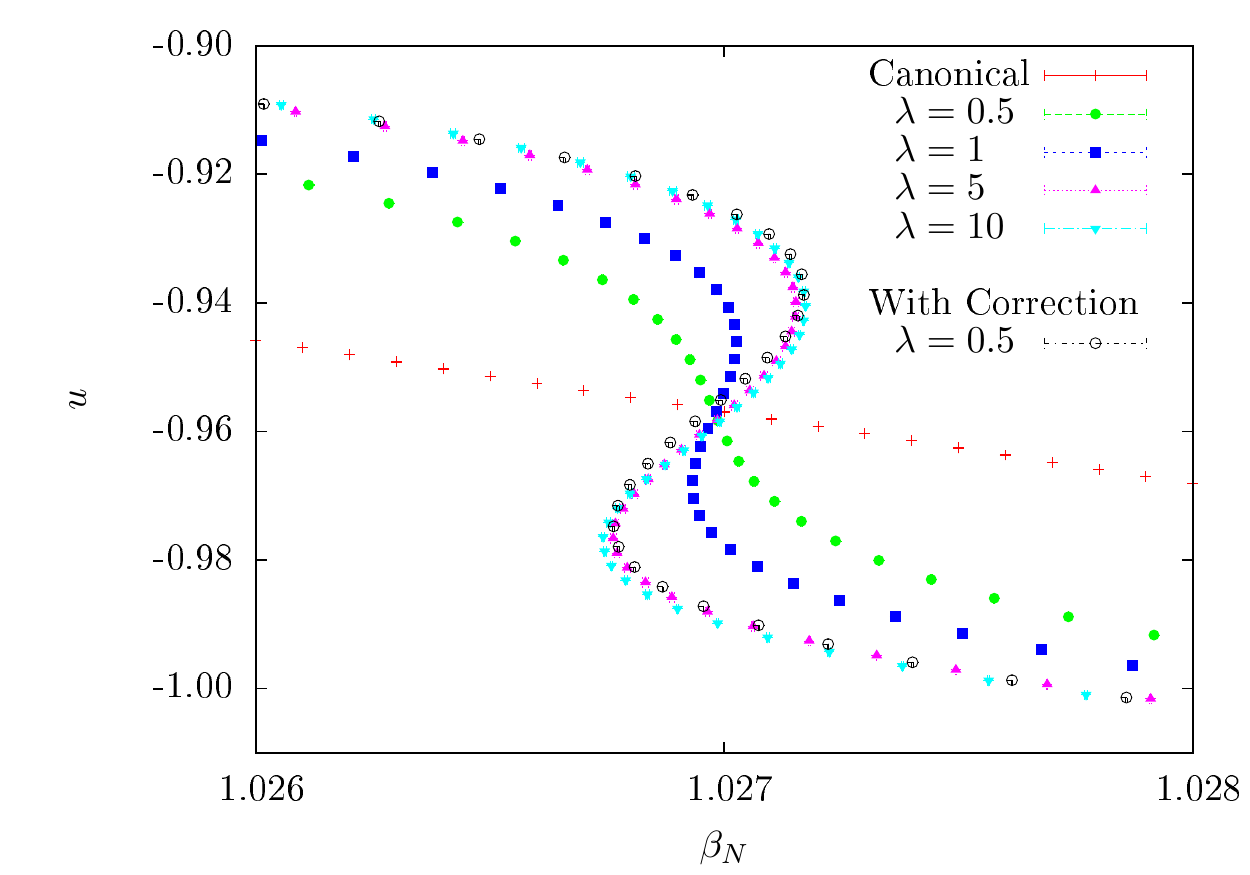}
  \caption{Relation between $\beta_N$ and $u$ for $L=64$,
 obtained using the canonical ensemble (crosses),
 and the squeezed ensembles with various values of $\lambda$ (filled symbols).
The result after the correction is also plotted for $\lambda=0.5$ (open circles).
}
\label{fig:fi_L64g060lambda}
\end{figure}

Fig.~\ref{fig:fi_L64g060lambda} shows the relation between $u$ and $\beta_N$ for $L=64$.
We calculate $\beta_N$ using the canonical ensemble and the squeezed ensembles
 with various values of  $\lambda$.
When the canonical ensemble is used, 
$u$ 
is given as a function of $\beta$,
which 
 is a single-valued function,  
 changing monotonically and continuously,
 even when $\sigma_N$ is not concave.
Hence, the effect due to the first-order phase transition is greatly diminished.
This should be contrasted with 
the results of the squeezed ensemble, 
which show that,
for sufficiently large $\lambda$,  
 $\beta_N$ becomes a multi-valued, S-shaped function in the transition region.
Consequently, 
one can correctly obtain the negative specific heat
 due to the first-order phase transition
by using the squeezed ensemble.

It is also seen from Fig.~\ref{fig:fi_L64g060lambda} that 
the functional form of $\beta_N(u)$ without correction (full circles)
becomes insensitive to the magnitude of $\lambda$ 
for sufficiently large $\lambda$.
This is because 
the error in $\beta_N$ by Eq.~(\ref{eq:temperature})
scales as $O(N^{-1}\lambda^{-1})$.
In fact, Eq.~(\ref{eq:u_high}) gives
\begin{align}
  \beta_N^\eta(\kappa)
  &\equiv \frac{\partial \eta}{\partial u} (\kappa, (u_N^\eta(\kappa))\\
  &= \beta_N (u_N^\eta(\kappa))
  + \frac{
    \frac{\partial^3 \xi_N^\eta}{\partial u^3} (\kappa, \upsilon_N^\eta(\kappa))
  }{
    2N \left|\frac{\partial^2 \xi_N^\eta}{\partial u^2} (\kappa, \upsilon_N^\eta(\kappa))\right|
  }
  + O(N^{-2})\\
  &=\beta_N (u_N^\eta(\kappa))
  + \frac{
    \sigma_N'''(\upsilon_N^\eta(\kappa))
  }{
    2N \lambda \left|1-\frac{\sigma_N''(\upsilon_N^\eta(\kappa))}{\lambda}\right|
  }
  + O(N^{-2}).
\end{align}
Therefore,
 the larger $\lambda$ gives the smaller error,
when using Eq.~(\ref{eq:temperature}).
On the other hand, the computational efficiency decreases
 with $\lambda$
 because the acceptance probability
 for configurations with higher energy
 is $e^{-O(N\lambda)}$.
To compromise these conflicting demands, 
we can take $\lambda$ small (to get a good efficiency) such as $\lambda=0.5$
and correct the result using the formulas derived in Section~\ref{sec:inequivalence}
(to decrease the error).
In fact, 
 we can obtain the relation between $u$ and $\beta_N$ with an error of $O(N^{-2})$
 from Eqs.~(\ref{eq:upsilon}) and (\ref{eq:u_high})
 using the squeezed ensemble with small $\lambda$
(open circles in Fig.~\ref{fig:fi_L64g060lambda}),
 and it agrees well with $\beta_N$ calculated using the squeezed ensemble with larger $\lambda$.

\begin{figure}
  \centering
  \includegraphics[width=1.0\linewidth]{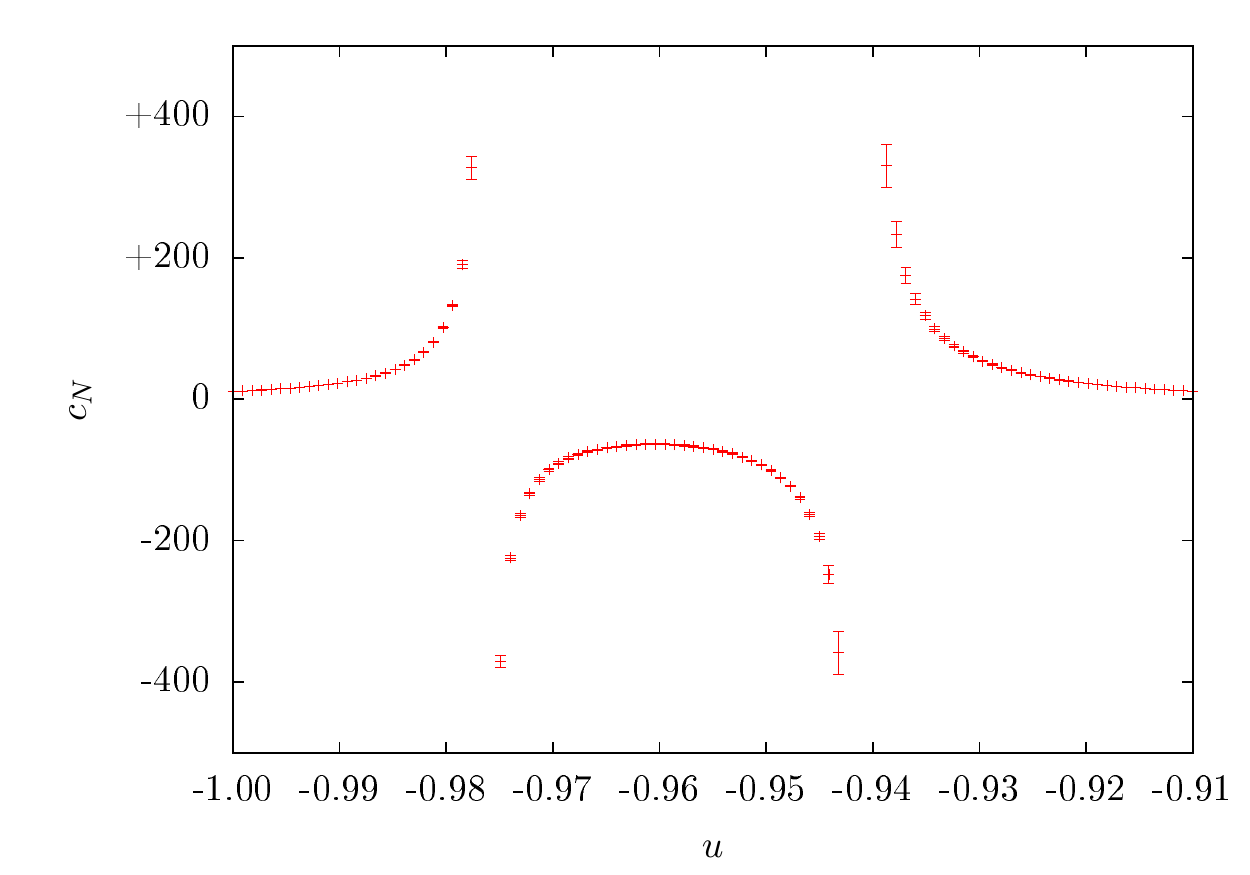}
  \caption{Relation between $c_N$ and $u$ for $L=64$ and $\lambda=5$.}
\label{fig:fi_L64g060lambda5_c}
\end{figure}

As discussed in Sec.~\ref{sec:non-concave},
 $c_N$ takes negative values in the transition region for finite $N$.
To confirm this fact,
 we calculate the relation between $u$ and $c_N$ for $L=64$
 from Eq.~(\ref{eq:specificheat}) using the squeezed ensemble with $\lambda=5$.
The results are plotted in Fig.~\ref{fig:fi_L64g060lambda5_c}.
It is confirmed that $c_N$ takes negative values between the two singular points.

\begin{figure}
  \centering
  \includegraphics[width=1.0\linewidth]{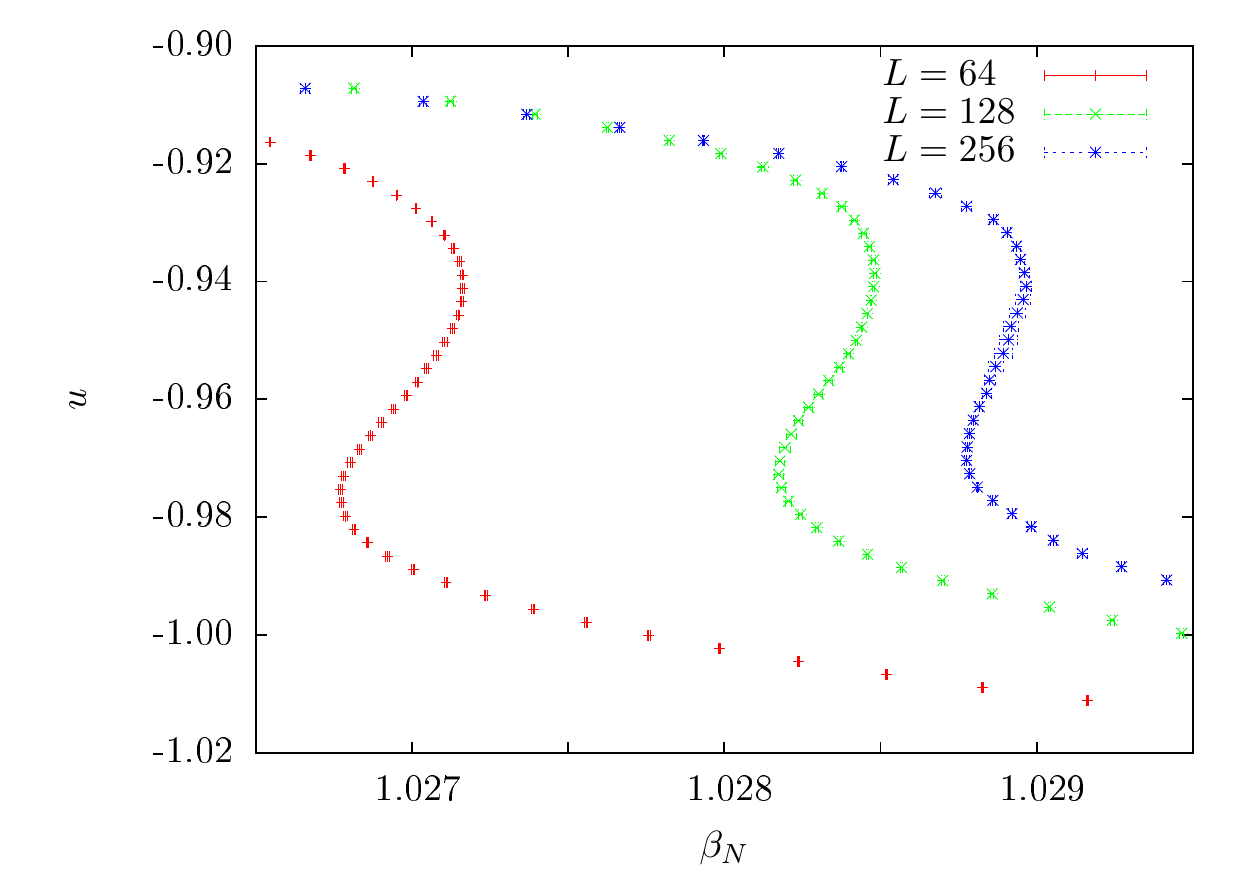}
  \caption{Relation between $\beta_N$ and $u$ for various values of $L$ and $\lambda=5$.}
\label{fig:fi_Lg060lambda5}
\end{figure}

Finally, we investigate how the result approaches
 that of the infinite system.
Fig.~\ref{fig:fi_Lg060lambda5} shows the relation
 between $\beta_N$ and $u$ for various values of $L$,
 where $\beta_N$ is calculated using the squeezed ensemble with $\lambda=5$.
With increasing $L$, the S-shaped region shifts towards the transition temperature of the infinite system,
 and this makes it possible to precisely extrapolate the transition temperature
 \cite{Challa1986,Borgs1992,Borgs1991,Challa1988_PRA}.
As $L$ increases, the amplitude of the S-shape decreases.
This implies that the concavity of $\sigma_N$ is broken due to the finite system size.

\section{Summary}
To summarize, we have proposed the squeezed ensembles (Sections~\ref{sec:SE}-\ref{sec:parameter}).
They give the correct equilibrium state
 even in the first-order phase transition region,
 as the microcanonical ensemble does.
In particular, for finite systems, 
one can correctly obtain thermodynamic anomalies
 such as negative specific heat,
 which appear generally in the transition region
  even when interactions are short-ranged
 (Sections~\ref{sec:non-concave}-\ref{sec:phys_nat_non-concave}
 and Appendix~\ref{sec:concav_breaking}).
Moreover, 
 the squeezed ensembles have good analytic properties, 
which yield useful analytic formulas (Section~\ref{sec:thermo_var})
 including the one by which temperature is obtained
 directly from energy without knowing entropy (Eq.~(\ref{eq:temperature})).
The squeezed ensembles are convenient for practical calculations
 because they can be numerically constructed easily,
 and efficient numerical methods, such as the replica exchange method,
 are applicable straightforwardly.
We also derive formulas which relate
 statistical-mechanical quantities of different ensembles
 for finite systems (Section~\ref{sec:inequivalence}).
The formulas are useful for obtaining results 
 with smaller finite-size effects,
 and for improving the computational efficiency.
We have confirmed the advantages of the squeezed ensembles and these formulas 
 by applying them to the Heisenberg model (Section~\ref{sec:Heisenberg})
 and the frustrated Ising model (Section~\ref{sec:FrustratedIsing}).

\begin{acknowledgments}
We thank Y. Chiba, K. Hukushima, H. Tasaki, H. Hakoshima, and S. Sugiura for discussions. 
This work was supported by The Japan Society
for the Promotion of Science, KAKENHI No. 15H05700 and No. 19H01810.
\end{acknowledgments} 

\appendix
\section{Proof of Eqs.~(\ref{eq:mech_var_exp})-(\ref{eq:mech_var_var})} \label{sec:proof_mech_var}
Consider systems with the translationally invariance.
We make a reasonable assumption that,
 by taking an appropriate constant $C$ depending only on $N$, 
 the expectation value of $\hat{a} \equiv \hat{A}/C$
 in the microcanonical ensemble
 converges to an $N$-independent continuous function of $u$ as $N\to\infty$. 
%
%
Then, by symmetry, the expectation value
 approaches a constant multiple of the expectation value of an appropriate additive observable
 as $N \to \infty$.

We denote by $\hat{\rho}_N^\mathrm{mc}(u-\delta(u), u]$
 the microcanonical ensemble with the energy shell $(u-\delta(u), u]$.
There is an arbitrariness in the choice of $\delta$.
Here, we take $\delta$ as follows.
Let $\delta_0$ be
 a positive constant of $O(N^0)$ such that
 $\epsilon^\mathrm{min}+\delta_0 < \upsilon_N^\eta$,
 and let $\delta$ be a positive valued function such that
\begin{align}
  \int_{u'-\delta(u')}^{u'} du e^{N\sigma_N(u)}
  = \int_{\epsilon^\mathrm{min}}^{\epsilon^\mathrm{min}+\delta_0} du e^{N\sigma_N(u)} \label{eq:delta}
\end{align}
 for all $u' \in [\epsilon^\mathrm{min}+\delta_0, \epsilon^\mathrm{max}]$.
Then it holds
\begin{align}
  \mathrm{Tr} \left[ \hat{a} \hat{\rho}_N^\eta \right]
  &= \int du \mathrm{Tr} \left[ \hat{a} \hat{\rho}_N^\mathrm{mc}(u-\delta(u), u] \right]
     \frac{e^{N \xi_N^\eta(u)}}{\Phi_N^\eta} + e^{-O(N)}, \label{eq:mc_decomp_1}\\
  \mathrm{Tr} \left[ \hat{a}^2 \hat{\rho}_N^\eta \right]
  &= \int du \mathrm{Tr} \left[ \hat{a}^2 \hat{\rho}_N^\mathrm{mc}(u-\delta(u), u] \right]
     \frac{e^{N \xi_N^\eta(u)}}{\Phi_N^\eta} + e^{-O(N)}, \label{eq:mc_decomp_2}
\end{align}

Using Eq.~(\ref{eq:mc_decomp_1}) and applying Laplace's method,
 we have
\begin{align}
  \mathrm{Tr} \left[ \hat{a} \hat{\rho}_N^\eta \right]
  &= \mathrm{Tr} \left[ \hat{a} \hat{\rho}_N^\mathrm{mc}(\upsilon_N^\eta-\delta(\upsilon_N^\eta), \upsilon_N^\eta] \right]
     + O(N^{-1})\\
  &= \mathrm{Tr} \left[ \hat{a} \hat{\rho}_N^\mathrm{mc}(u_N^\eta) \right] + O(N^{-1}).
\end{align}
Therefore, we have Eq.~(\ref{eq:mech_var_exp}).

Since we assumed that the equilibrium state is specified by the energy density,
 we have
\begin{align}
  \mathrm{Tr}\left[\hat{a}^2 \hat{\rho}_N^\mathrm{mc}(u) \right]
  -\mathrm{Tr}\left[\hat{a} \hat{\rho}_N^\mathrm{mc}(u) \right]^2
  = o(N^0)
\end{align}
 for all macroscopic additive observable $\hat{A}$.
Then using Eqs.~(\ref{eq:mc_decomp_2}) and applying Laplace's method,
 we have
\begin{widetext}
\begin{align}
  &\mathrm{Tr}\left[\hat{a}^2 \hat{\rho}_N^\eta\right]
  - \mathrm{Tr}\left[\hat{a} \hat{\rho}_N^\eta\right]^2\\
  &= \int du \underbrace{\left(
    \mathrm{Tr}\left[\hat{a}^2 \hat{\rho}_N^\mathrm{mc}(u-\delta(u), u]\right]
   -\mathrm{Tr}\left[\hat{a} \hat{\rho}_N^\mathrm{mc}(u-\delta(u), u]\right]^2
  \right)}_{o(N^0)} \frac{e^{N \xi_N^\eta(u)}}{\Phi_N^\eta}\\
  &+ \underbrace{
       \int du \mathrm{Tr}\left[\hat{a} \hat{\rho}_N^\mathrm{mc}(u-\delta(u), u]\right]^2
       \frac{e^{N \xi_N^\eta(u)}}{\Phi_N^\eta}
     - \left( \int du \mathrm{Tr}\left[\hat{a} \hat{\rho}_N^\mathrm{mc}(u-\delta(u), u]\right]
       \frac{e^{N \xi_N^\eta(u)}}{\Phi_N^\eta} \right)^2
     }_{O(N^{-1})}
   + e^{-O(N)}\\
  &= o(N^0),
\end{align}
\end{widetext}
 for all macroscopic additive observable $\hat{A}$.
Therefore, we have Eq.~(\ref{eq:mech_var_var}).

\section{Proof of Eqs.~(\ref{eq:psi-s})-(\ref{eq:s-psi})} \label{sec:proof_equivalence}
Eq.~(\ref{eq:s-psi})
 can be immediately obtained from Eq.~(\ref{eq:massieu}).
Eq.~(\ref{eq:psi-s}) is proved as follows.
Using Eq.~(\ref{eq:s-psi}), we have
\begin{align}
  s(u)
  &\leq \eta(\kappa, u) - \inf_u \left\{ \eta(\kappa, u)-s(u) \right\} \\
  &= \eta(\kappa, u) - \psi^\eta(\kappa)
\end{align}
 for all $\kappa$.
Then we have
\begin{align}
   s(u) \leq \inf_\kappa \left\{ \eta(\kappa, u)-\psi^\eta(\kappa) \right\}.
\end{align}
On the other hand, using Eq.~(\ref{eq:s-psi}), we have
\begin{align}
  &\inf_\kappa \left\{ \eta(\kappa, u)-\psi^\eta(\kappa) \right\}\\
  &\leq \eta(\kappa, u)-\psi^\eta(\kappa)\\
  &= \eta(\kappa, u) - \inf_u \left\{ \eta(\kappa, u)-s(u) \right\}\\
  &= \eta(\kappa, u) - \eta(\kappa, \upsilon_\infty^\eta(\kappa)) + s(\upsilon_\infty^\eta(\kappa))
\end{align}
 for all $\kappa$.
The continuity of $\upsilon_\infty^\eta$ in $\kappa$ (Eq.~(\ref{eq:upsilon_p})) implies that there exists $\kappa$ such that $\upsilon_\infty^\eta(\kappa)=u$, then it holds
\begin{align}
  \inf_\kappa \left\{ \eta(\kappa, u)-\psi^\eta(\kappa) \right\} &\leq s(u).
\end{align}
Therefore we have Eq.~(\ref{eq:psi-s}).

\section{Generality of Concavity breaking of $\sigma_N$} \label{sec:concav_breaking}
We prove that the concavity breaking {\em always} occurs for finite $N$
 under reasonable conditions.
For proper handling of a system which undergoes a first-order phase transition,
 we consider a more general case where the equilibrium state is specified using a general set of extensive variables.

\subsection{Concavity breaking in liquid-gas systems} \label{subsec:concav_breaking_liquid-gas}
As an example, we take a $d$-dimensional system which undergoes a first-order liquid-gas phase transition
 (whereas general systems will be discussed in Appendix~\ref{subsec:concav_breaking_gen}).
Its equilibrium state is assumed to be specified by the energy $U$, volume $V$ and number of particles $N$.
Hence, for a given value of $N$, $\sigma_N$ is
 a function of the energy per particle, $u \equiv U/N$, and the volume per particle, $v \equiv V/N$, i,e.,
\begin{align}
  \sigma_N = \sigma_N(u,v).
\end{align}

We investigate properties of $\sigma_N(u,v)$, for each fixed value of $N$, in the state space spanned by $u$ and $v$.
Note that in this state space the liquid-gas coexisting region is a two-dimensional region,
 whereas it is a one-dimensional region (line) in the state space spanned by temperature $T$ and pressure $P$.

Suppose that the system is in an equilibrium state where liquid and gas phases coexist.
We assume that
 the thickness and surface area of phase boundaries are $O(N^0)$ and $O(N^{1-1/d})$, respectively.
Furthermore, we assume that finite size effects in each phase are smaller than
 those due to the phase boundaries.
This condition is natural and reasonable, as discussed in Appendix~\ref{sec:concav_condition}.
Under the above conditions,
 we show that the concavity of $\sigma_N$ is {\em always} broken for finite $N$ in such an equilibrium state.

\begin{figure*}[t]
  \begin{subfigure}[t]{0.49\linewidth}
    \centering
  \begin{tikzpicture}
    \draw[decorate, decoration={brace,mirror}, yshift=-1ex]  (0\paperwidth,-0.06\paperwidth)
    -- node[below=0.4ex] {$N$}  (0.35\paperwidth,-0.06\paperwidth);
    \path (0,0) node[rectangle,
    anchor=west,
    fill=white!50!blue,
    minimum width=0.22\paperwidth, minimum height=0.12\paperwidth,
    align=center]{\Huge Liquid\\ \\$(\tilde{u}^\mathrm{Liquid},\tilde{v}^\mathrm{Liquid})$}
    -- (0.22\paperwidth,0) node[rectangle,
    anchor=west,
    fill=white!50!red,
    minimum width=0.13\paperwidth, minimum height=0.12\paperwidth,
    align=center]{\Huge Gas\\ \\$(\tilde{u}^\mathrm{Gas},\tilde{v}^\mathrm{Gas})$};
    \draw[thick] (0\paperwidth,-0.06\paperwidth)
    -- (0.35\paperwidth,-0.06\paperwidth)
    -- (0.35\paperwidth,+0.06\paperwidth)
    -- (0\paperwidth,+0.06\paperwidth)
    -- (0\paperwidth,-0.06\paperwidth);
  \end{tikzpicture}
  \subcaption{} \label{fig:N_PB_can_neg}
  \end{subfigure}
  \begin{subfigure}[t]{0.49\linewidth}
    \centering
  \begin{tikzpicture}
    \draw[decorate, decoration={brace,mirror}, yshift=-1ex]  (0\paperwidth,-0.06\paperwidth)
    -- node[below=0.4ex] {$N$}  (0.35\paperwidth,-0.06\paperwidth);
    \path (0,0) node[rectangle,
    anchor=west,
    fill=white!30!blue,
    minimum width=0.17\paperwidth, minimum height=0.12\paperwidth,
    align=center]{\Huge Liquid\\ \\$(u^\mathrm{Liquid},v^\mathrm{Liquid})$}
    -- (0.17\paperwidth,0) node[
    rectangle,
    anchor=west,
    left color=white!30!blue, right color=white!30!red,
    shading angle=90,
    minimum width=0.05\paperwidth, minimum height=0.12\paperwidth]{}
    -- (0.22\paperwidth,0) node[rectangle,
    anchor=west,
    fill=white!30!red,
    minimum width=0.13\paperwidth, minimum height=0.12\paperwidth,
    align=center]{\Huge Gas\\ \\$(u^\mathrm{Gas},v^\mathrm{Gas})$};
    \draw[thick] (0\paperwidth,-0.06\paperwidth)
    -- (0.35\paperwidth,-0.06\paperwidth)
    -- (0.35\paperwidth,+0.06\paperwidth)
    -- (0\paperwidth,+0.06\paperwidth)
    -- (0\paperwidth,-0.06\paperwidth);
    \draw[dotted, thick] (0.17\paperwidth,-0.06\paperwidth)--(0.17\paperwidth,+0.06\paperwidth);
    \draw[dotted, thick] (0.22\paperwidth,-0.06\paperwidth)--(0.22\paperwidth,+0.06\paperwidth);
  \end{tikzpicture}
  \subcaption{} \label{fig:N_PB_cannot_neg}
  \end{subfigure}
  \caption{Comparison of the equilibrium states with the same values of $(u,v)$ for two cases
   where \subref{fig:N_PB_can_neg} the effects of the phase boundaries are ignored and \subref{fig:N_PB_cannot_neg} not.}
  \label{fig:N_PhaseBoundary}
\end{figure*}
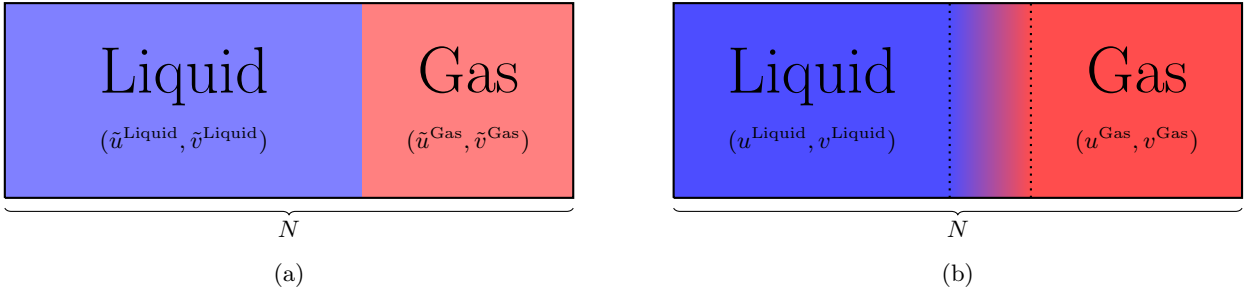

Let us compare the equilibrium states with the same values of $(u, v)$ for two cases
 where
 \subref{fig:N_PB_can_neg} the effects of the phase boundaries are ignored
 and
 \subref{fig:N_PB_cannot_neg} not (Fig.~\ref{fig:N_PhaseBoundary}).

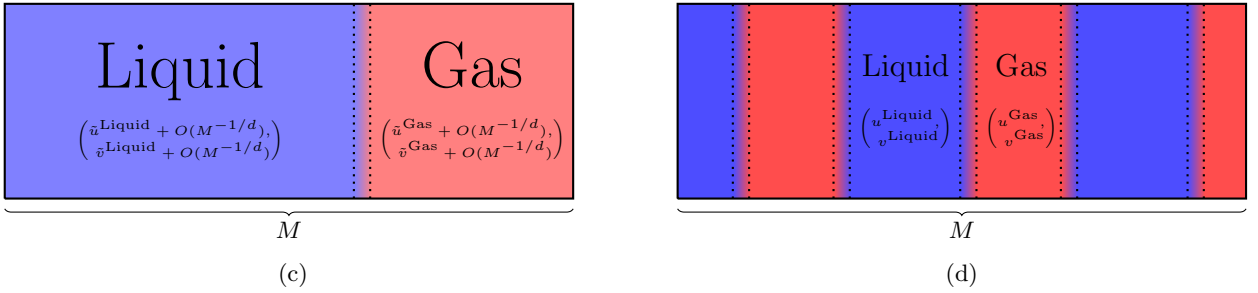
\begin{figure*}
  \begin{subfigure}[t]{0.49\linewidth}
    \centering
  \begin{tikzpicture}
    \draw[decorate, decoration={brace,mirror}, yshift=-1ex]  (0\paperwidth,-0.06\paperwidth)
    -- node[below=0.4ex] {$M$}  (0.35\paperwidth,-0.06\paperwidth);
    \path (0,0) node[rectangle,
    anchor=west,
    fill=white!50!blue,
    minimum width=0.215\paperwidth, minimum height=0.12\paperwidth,
    align=center]{\Huge Liquid\\ \\
    \tiny
    $\left(
      \substack{\displaystyle\tilde{u}^\mathrm{Liquid}+O(M^{-1/d}),\\
              \ \displaystyle\tilde{v}^\mathrm{Liquid}+O(M^{-1/d})}
    \right)$}
    -- (0.215\paperwidth,0) node[
    rectangle,
    anchor=west,
    left color=white!50!blue, right color=white!50!red,
    shading angle=90,
    minimum width=0.01\paperwidth, minimum height=0.12\paperwidth]{}
    -- (0.225\paperwidth,0) node[rectangle,
    anchor=west,
    fill=white!50!red,
    minimum width=0.125\paperwidth, minimum height=0.12\paperwidth,
    align=center]{\Huge Gas\\ \\
    \tiny
    $\left(
      \substack{\displaystyle\tilde{u}^\mathrm{Gas}+O(M^{-1/d}),\\
              \ \displaystyle\tilde{v}^\mathrm{Gas}+O(M^{-1/d})}
    \right)$};
    \draw[thick] (0\paperwidth,-0.06\paperwidth)
    -- (0.35\paperwidth,-0.06\paperwidth)
    -- (0.35\paperwidth,+0.06\paperwidth)
    -- (0\paperwidth,+0.06\paperwidth)
    -- (0\paperwidth,-0.06\paperwidth);
    \draw[dotted, thick] (0.215\paperwidth,-0.06\paperwidth)--(0.215\paperwidth,+0.06\paperwidth);
    \draw[dotted, thick] (0.225\paperwidth,-0.06\paperwidth)--(0.225\paperwidth,+0.06\paperwidth);
  \end{tikzpicture}
  \setcounter{subfigure}{2}
  \subcaption{} \label{fig:M_PB_can_neg}
  \end{subfigure}
  \begin{subfigure}[t]{0.49\linewidth}
    \centering
  \begin{tikzpicture}
    \draw[decorate, decoration={brace,mirror}, yshift=-1ex]  (0\paperwidth,-0.06\paperwidth)
    -- node[below=0.4ex] {$M$}  (0.35\paperwidth,-0.06\paperwidth);
    \path (0,0) node[rectangle,
    anchor=west,
    fill=white!30!blue,
    minimum width=0.034\paperwidth, minimum height=0.12\paperwidth,
    align=center]{}
    -- (0.034\paperwidth,0) node[
    rectangle,
    anchor=west,
    left color=white!30!blue, right color=white!30!red,
    shading angle=90,
    minimum width=0.01\paperwidth, minimum height=0.12\paperwidth]{}
    -- (0.044\paperwidth,0) node[rectangle,
    anchor=west,
    fill=white!30!red,
    minimum width=0.052\paperwidth, minimum height=0.12\paperwidth,
    align=center]{}
    -- (0.096\paperwidth,0) node[
    rectangle,
    anchor=west,
    left color=white!30!red, right color=white!30!blue,
    shading angle=90,
    minimum width=0.01\paperwidth, minimum height=0.12\paperwidth]{}
    -- (0.106\paperwidth,0) node[rectangle,
    anchor=west,
    fill=white!30!blue,
    minimum width=0.068\paperwidth, minimum height=0.12\paperwidth,
    align=center]{\large Liquid\\ \\
    \tiny
    $\left(
      \substack{\displaystyle u^\mathrm{Liquid},\\
              \ \displaystyle v^\mathrm{Liquid}}
    \right)$}
    -- (0.174\paperwidth,0) node[
    rectangle,
    anchor=west,
    left color=white!30!blue, right color=white!30!red,
    shading angle=90,
    minimum width=0.01\paperwidth, minimum height=0.12\paperwidth]{}
    -- (0.184\paperwidth,0) node[rectangle,
    anchor=west,
    fill=white!30!red,
    minimum width=0.052\paperwidth, minimum height=0.12\paperwidth,
    align=center]{\large Gas\\ \\
    \tiny
    $\left(
      \substack{\displaystyle u^\mathrm{Gas},\\
              \ \displaystyle v^\mathrm{Gas}}
    \right)$}
    -- (0.236\paperwidth,0) node[
    rectangle,
    anchor=west,
    left color=white!30!red, right color=white!30!blue,
    shading angle=90,
    minimum width=0.01\paperwidth, minimum height=0.12\paperwidth]{}
    -- (0.246\paperwidth,0) node[rectangle,
    anchor=west,
    fill=white!30!blue,
    minimum width=0.068\paperwidth, minimum height=0.12\paperwidth,
    align=center]{}
    -- (0.314\paperwidth,0) node[
    rectangle,
    anchor=west,
    left color=white!30!blue, right color=white!30!red,
    shading angle=90,
    minimum width=0.01\paperwidth, minimum height=0.12\paperwidth]{}
    -- (0.324\paperwidth,0) node[rectangle,
    anchor=west,
    fill=white!30!red,
    minimum width=0.026\paperwidth, minimum height=0.12\paperwidth,
    align=center]{};
    \draw[thick] (0\paperwidth,-0.06\paperwidth)
    -- (0.35\paperwidth,-0.06\paperwidth)
    -- (0.35\paperwidth,+0.06\paperwidth)
    -- (0\paperwidth,+0.06\paperwidth)
    -- (0\paperwidth,-0.06\paperwidth);
    \draw[dotted, thick] (0.034\paperwidth,-0.06\paperwidth)--(0.034\paperwidth,+0.06\paperwidth);
    \draw[dotted, thick] (0.044\paperwidth,-0.06\paperwidth)--(0.044\paperwidth,+0.06\paperwidth);
    \draw[dotted, thick] (0.096\paperwidth,-0.06\paperwidth)--(0.096\paperwidth,+0.06\paperwidth);
    \draw[dotted, thick] (0.106\paperwidth,-0.06\paperwidth)--(0.106\paperwidth,+0.06\paperwidth);
    \draw[dotted, thick] (0.174\paperwidth,-0.06\paperwidth)--(0.174\paperwidth,+0.06\paperwidth);
    \draw[dotted, thick] (0.184\paperwidth,-0.06\paperwidth)--(0.184\paperwidth,+0.06\paperwidth);
    \draw[dotted, thick] (0.236\paperwidth,-0.06\paperwidth)--(0.236\paperwidth,+0.06\paperwidth);
    \draw[dotted, thick] (0.246\paperwidth,-0.06\paperwidth)--(0.246\paperwidth,+0.06\paperwidth);
    \draw[dotted, thick] (0.314\paperwidth,-0.06\paperwidth)--(0.314\paperwidth,+0.06\paperwidth);
    \draw[dotted, thick] (0.324\paperwidth,-0.06\paperwidth)--(0.324\paperwidth,+0.06\paperwidth);
  \end{tikzpicture}
  \subcaption{} \label{fig:M_PB_cannot_neg}
  \end{subfigure}
  \caption{Comparison of the microstates with sufficiently large $M$ and the same values of $(u, v)$ for two cases
   where \subref{fig:M_PB_can_neg} surface area of the phase boundaries is $O(M^{1-1/d})$
   and \subref{fig:M_PB_cannot_neg} $M/N$ times that in the equilibrium state with the number of particles $N$.}
  \label{fig:M_PhaseBoundary}
\end{figure*}

First, we consider case \subref{fig:N_PB_can_neg}.
We denote a physical quantity in this case by a tilde over its symbol, such as $\tilde{\sigma}_N$.
Since phase boundaries are ignored,
 one can define quantities (such as the number of particles) in each phase without ambiguity.
Let $\tilde{N}^p$, $\tilde{u}^p$, and $\tilde{v}^p$
 be the number of particles, energy per particle, and volume per particle, respectively, in phase $p$.
Since phase boundaries are ignored, the fraction $\tilde{\nu}^p \equiv \tilde{N}^p/N$ satisfies
\begin{align}
  \sum_p\tilde{\nu}^p = 1, \label{eq:tilde_nu}
\end{align}
 and $u$, $v$ and $\tilde{\sigma}_N$
 agree with the weighted arithmetic means as
\begin{align}
  u &= \sum_p\tilde{\nu}^p\tilde{u}^p, \label{eq:tilde_u}\\
  v &= \sum_p\tilde{\nu}^p\tilde{v}^p, \label{eq:tilde_v}\\
  \tilde{\sigma}_N(u, v) &= \sum_p\tilde{\nu}^p \sigma_N(\tilde{u}^p, \tilde{v}^p). \label{eq:tilde_sigma}
\end{align}

Next, we consider case \subref{fig:N_PB_cannot_neg} where effects of phase boundaries are not ignored.
For $U, V$ and $N$ of the whole system including the phase boundaries,
 we take their values same as those in case \subref{fig:N_PB_can_neg}.
Then
 the fraction and the state of the bulk of each phase change slightly (if possible)
 as compared with the case where the phase boundaries can be neglected.
Hence, compared with $\tilde{\sigma}_N$,
 the $\sigma_N$ changes according to two main causes:
 the number of possible configurations of the phase boundaries
 and the changes in the fraction and the state of the bulk of each phase.

Let us compare $\sigma_N(u,v)$ and $\tilde{\sigma}_N(u, v)$.
For this purpose,
 we consider an equilibrium state with the number of particles $M$ and the same values of $(u, v)$,
 where $1 \ll M/N$.
Let us compare the microstates for two cases
 where
 \subref{fig:M_PB_can_neg} surface area of the phase boundaries is $O(M^{1-1/d})$
 and
 \subref{fig:M_PB_cannot_neg} $M/N$ times that in the equilibrium state with the number of particles $N$ (Fig.~\ref{fig:M_PhaseBoundary}).
The density of microstates \subref{fig:M_PB_can_neg}
 agree with $e^{M \left( \tilde{\sigma}_N(u, v)+O(M^{-1/d}) \right)}$.
On the other hand, the density of microstates \subref{fig:M_PB_cannot_neg}
 is larger than $e^{M \sigma_N(u, v)}$.
Therefore, If $\sigma_N(u, v)$ were larger than $\tilde{\sigma}_N(u, v)$,
 in the sufficiently large system,
 the density of microstates \subref{fig:M_PB_can_neg}
 would be exponentially smaller
 than the density of microstates \subref{fig:M_PB_cannot_neg}.
Hence, the macrostate with the phase boundaries
 whose surface area is $O(M^{1-1/d})$
 would not be realized as an equilibrium state which is a typical macrostate with the largest number of microstates.
Since such an equilibrium state contradicts with our assumption,
 we conclude
\begin{align}
  \sigma_N(u, v) < \tilde{\sigma}_N(u, v) \label{eq:sigma―tilde_sigma}
\end{align}
From Eqs.~(\ref{eq:tilde_nu})-(\ref{eq:sigma―tilde_sigma}), we find
\begin{widetext}
\begin{align}
  \sigma_N \left(
             \tilde{\nu}^\mathrm{Gas}         \left( \tilde{u}^\mathrm{Gas},    \tilde{v}^\mathrm{Gas}    \right)
  + \left( 1-\tilde{\nu}^\mathrm{Gas} \right) \left( \tilde{u}^\mathrm{Liquid}, \tilde{v}^\mathrm{Liquid} \right)
  \right)
  <          \tilde{\nu}^\mathrm{Gas}         \sigma_N \left( \tilde{u}^\mathrm{Gas},    \tilde{v}^\mathrm{Gas}    \right)
  + \left( 1-\tilde{\nu}^\mathrm{Gas} \right) \sigma_N \left( \tilde{u}^\mathrm{Liquid}, \tilde{v}^\mathrm{Liquid} \right)
\end{align}
\end{widetext}
 for certain values of $\tilde{u}^p, \tilde{v}^p$ and $\tilde{\nu}^\mathrm{Gas} \in (0, 1)$.
This shows that the concavity of $\sigma_N(u,v)$ is broken.

This concavity breaking occurs in equilibrium states where liquid and gas phases coexist, i.e.,
 in the phase-coexisting region in the state space spanned by $(u,v)$.

\subsection{Concavity breaking in general systems} \label{subsec:concav_breaking_gen}
The above discussions can easily be extended to general systems
 with short-range interactions, as follows.

Consider a $d$-dimensional system whose equilibrium states is specified
 by extensive variables $X_1, X_2, \cdots, X_M$ and $N$.
Hence, for a given value of $N$,
 $\sigma_N$ is a function of $(x_1, x_2, \cdots, x_M)$, where $x_i \equiv X_i/N$.
Suppose that the system which undergoes a first-order phase transition
 and that several phases coexist in the phase transition region.
We assume that
 the thickness and surface area of phase boundaries are $O(N^0)$ and $O(N^{1-1/d})$, respectively.
Furthermore, we assume that finite size effects in each phase are smaller than
 those due to the phase boundaries.
Then concavity of $\sigma_N$ is {\em always} broken for finite $N$.

\subsection{Concavity restoration in the thermodynamic limit}
The anomalous behavior of $\sigma_N$ is peculiar to the finite systems
 and the concavity of $\sigma_N$ recovers in the thermodynamic limit\cite{Lynden-Bell1995,Ispolatov2001}.
Since the surface area of the phase boundaries is $O(N^{1-1/d})$.
 the contribution of the phase boundaries to $x_i$ and $\sigma_N$ is $O(N^{-1/d})$.
On the other hand, in the model with the long-range interaction obtained using the mean-field approximation,
 the concavity of $\sigma_N$ is broken even in the thermodynamic limit.
Therefore, an S-shaped caloric curve of such the model (so-called van der Waals loop)
 is an unphysical artifact of the approximation
 and must be distinguished from S-shaped $\beta_N$ in the finite system with short-range interaction.

\subsection{Validity of the condition} \label{sec:concav_condition}
In Appendices~\ref{subsec:concav_breaking_liquid-gas}-\ref{subsec:concav_breaking_gen},
 we have assumed that finite size effects can be neglected
 except for those due to the phase boundaries.
Here we discuss the validity of this condition.

Let us compare the asymptotic behavior of
 the finite size effects due to the phase boundaries and those due to other causes at large $N$.
We consider the case where surface effects can be neglected.

As mentioned in Appendix~\ref{subsec:concav_breaking_gen},
 the finite size effects due to the phase boundaries on $\sigma_N$ are $O(N^{-1/d})$.
With short-range interactions, the spatial dimension of the system
 which undergoes a first-order phase transition at finite temperature
 satisfies $d>1$\cite{Araki1969,Araki1975}.
On the other hand,
 the finite size effects excluding those due to the phase boundaries on $\sigma_N$
 is expected to be $O(N^{-1})$.
In fact, the difference
 between the transition points in finite and infinite systems
 is $O(N^{-1})$\cite{Challa1986}.

Therefore, for sufficiently large $N$,
 finite size effects can be neglected except for those due to the phase boundaries.

\section{Inapplicability of the canonical ensemble} \label{sec:non-concave_canonical}
We examine whether the canonical ensemble
 is applicable to the system with non-concave $\sigma_N$
 (Fig.~\ref{subfig:nc_sigma}).
The canonical ensemble corresponds to the case where $\eta = \beta u$,
 and the equilibrium state is specified by $\beta$.
Since
 neither condition~\ref{cond:A} nor \ref{cond:F}-\ref{cond:G}
 are satisfied by this choice of $\eta$, 
 the following difficulty arises for systems with non-concave $\sigma_N$.

If $\sigma_N$ were strictly convex,
 the energy density distribution in the canonical ensemble
 would have a single peak.
The peak position $\upsilon_N^\mathrm{can}$
 would be uniquely determined as the solution to $\beta_N(\upsilon_N^\mathrm{can})=\beta$
 (Eq.~(\ref{eq:upsilon})).
However, this property is lost in the case
 where $\sigma_N$ is not concave.

\begin{figure}
  \centering
  \includegraphics[keepaspectratio, width=\linewidth]{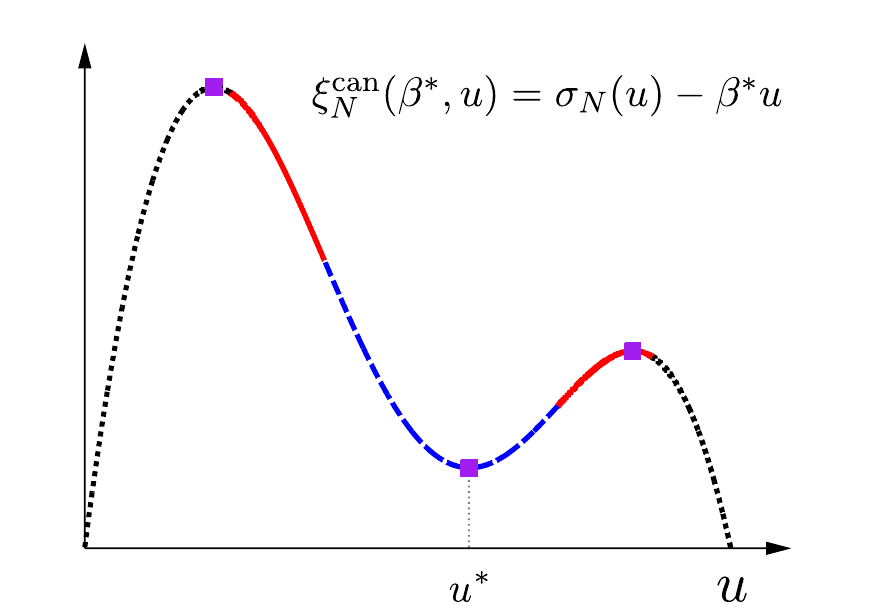}
  \caption{The energy density distribution in the canonical ensemble in the case where $\sigma_N$ is not concave.}
  \label{fig:non-concave_can}
\end{figure}

For example, suppose that one wants to investigate
 microscopic structures of the equilibrium state
 with the energy density $u^*$ on the blue dashed line
 in Fig.~\ref{fig:non-concave}.
This is impossible if one uses the canonical ensemble.
In fact,
 if one takes $\beta = \beta^* \equiv \beta_N(u^*)$ in the canonical ensemble,
 there are three points of $u$ 
 (purple squares in Fig.~\ref{subfig:nc_beta})
that give the same value of $\beta_N$.
Consequently,
 the energy distribution becomes bimodal, as shown in Fig.~\ref{fig:non-concave_can},
 and takes the {\em local minimum} at $u^*$.
Hence the desired equilibrium state is not obtained,
 and one cannot investigate its microscopic structures.
This should be contrasted with a squeezed ensemble,
 whose energy distribution has the {\em global maximum} at $u=u^*$,
 giving the desired equilibrium state correctly.

Similarly, for $u$ on the red solid line,
 the energy density distribution in the canonical ensemble at $\beta=\beta_N(u)$
 has the local (but not global) maximum at $u$.

To sum up,
 the equilibrium states with the energy density
 in the region other than the black dotted line
 in Fig.~\ref{fig:non-concave}
 cannot be obtained using the canonical ensemble.

\end{document}